\documentclass[seceq]{ptptex}

\usepackage{amsmath,amsfonts,latexsym,amssymb} 
\usepackage{verbatim,amsthm,graphicx} 
\usepackage{wrapft}
\usepackage{mathrsfs}

\def\ben{\begin{equation}}
\def\een{\end{equation}}
\def\bena{\begin{eqnarray}}
\def\eena{\end{eqnarray}}

\newcommand{\non}{\nonumber}
\theoremstyle{definition}

\newcommand{\I}{{\mathscr{I}}}

\newcommand{\bomega}{{\underline \Omega}}

\newcommand{\n}{{\underline m}}
\newcommand{\mr}{{\boldsymbol R}}
\newcommand{\mn}{{\boldsymbol N}}

\newcommand{\mz}{{\boldsymbol Z}}

\title{
Topology and Uniqueness of Higher Dimensional Black Holes  
}

\author{
Daisuke \textsc{Ida}$^{a}$, %
Akihiro \textsc{Ishibashi}$^{b c}$ and %
Tetsuya \textsc{Shiromizu}$^{d}$ %
}

\inst{
${}^{a}$Department of Physics, Gakushuin University, 
\\ 
Tokyo 171-8588, Japan
\\
${}^{b}$Theory Center, Institute of Particle and Nuclear Studies, 
\\
High Energy Accelerator Research Organization (KEK), 
\\
Tsukuba, 305-0801, Japan   
\\
$^{c}$Department of Physics, Kinki University, 
\\ 
Higashi-Osaka 577-8502, Japan
\\
and 
\\
${}^{d}$Department of Physics, Kyoto University, Kyoto 606-8502, Japan
}

\abst{
We review recent results concerning general properties of higher dimensional 
black holes. The topics selected with particular focus are those concerning 
topology, symmetry, and uniqueness properties of asymptotically flat vacuum 
black holes in higher dimensional general relativity. 
}

\begin{document}

\maketitle 

\tableofcontents 

\section{Introduction}
\label{sec:introduction}
The black hole uniqueness theorem in $4$-dimensions is a triumph 
of classical general relativity, implying that a tremendous 
number of black holes existing in our observable universe can 
be described accurately by the Kerr metric, which possesses 
only two parameters.  
In the course of complete proof of the uniqueness theorem,  
there have appeared a number of remarkable results, each of which 
itself reveals physically an important property of black holes, 
such as those concerning topology and symmetry\cite{H72}. 
These results also give us deep insights into thermodynamic aspects 
of black holes\cite{Wald01LR}.

It is of great interest to consider generalizations of a number of 
theorems established for $4$-dimensional black holes 
to higher dimensional case. 
One might expect that such a generalization could straightforwardly be done 
by merely replacing ``$4$'' with a general number ``$D$.'' 
However, as is by now well-known, the discovery of the black ring solution 
in $5$-dimensions~\cite{ER02b} (as well as a large variety of 
exact solutions discussed in Chapter~1 and~4) has drastically changed 
our view of the issue, highlighting that the uniqueness theorem 
no longer holds as it stands in higher dimensions.

In this chapter we shall review general properties of higher dimensional 
black holes, attempting to clarify which properties of $4$-dimensional 
black holes can be straightforwardly generalized to higher dimensions 
and which properties hold only in $4$-dimensions. 
There have already been a number of theorems for higher 
dimensional black holes. 
This review is, however, not intended to cover the whole relevant 
subjects or supply a complete list of existing literature. 
Rather, we will focus on some specific topics relevant to topology and 
uniqueness/non-uniqueness feature of asymptotically flat 
vacuum black holes and describe some key ideas and methods 
for obtaining the results in higher dimensions.

In the next section we shall first recapitulate how to describe 
global structure of black hole spacetimes in general dimensions.
Then we discuss topological aspects of apparent and the event horizon
of dynamical black holes. 
In section~\ref{sec:stationary}, 
we consider stationary black holes in higher dimensional 
general relativity. After briefly summarizing a few critical steps 
in the proof of black hole uniqueness theorems in $4$-dimensions, 
we shall discuss how and to what extent each of the steps can be 
generalized to higher dimensional setting. In particular, 
we review recent results concerning possible restrictions on the 
horizon topology and also an enhancement of Killing symmetries, 
called the rigidity theorem, in higher dimensions. 
In section~\ref{sec:unique:static}, we review uniqueness theorems 
for static black holes in higher dimensions. 
Uniqueness theorems for asymptotically flat, stationary 
rotating, vacuum black holes are discussed in detail 
in section~\ref{sec:unique:stationary}. 
Section~\ref{sec:summary} is devoted to summary. 

\medskip
\noindent
{\em Notations and Conventions}\\

In this chapter, we mainly use the abstract index notation for
tensor fields on a spacetime, where each slot for tangent or cotangent field is denoted by a lower-case latin index:
$a,b,c,\cdots$.
In section~\ref{sec:unique:static},
a tensor field is written in terms of its components, where
upper-case latin indices $M,N,\cdots$ run from $0$ to $D-1$,
and lower-case latin indices $i,j,\cdots$ run from $1$ to $D-1$.
In section~\ref{sec:unique:stationary},
 we mainly treat the $5$-dimensional spacetimes.
There, a tensor field is mainly written in terms of its components,
where lower-case latin indices $i,j,\cdots$ run from $1$ to $2$
and upper-case latin indices $I,J,\cdots$ run from $3$ to $5$.

We use the natural unit where the speed of light $c$ and the reduced
Planck constant $\hbar$ are set to unity.

\section{General properties of higher dimensional black holes}  
\label{sec:general}
We begin with noting that the uniqueness feature of a stationary black hole   
is related to its thermodynamic aspects in the sense that a 
thermodynamically equilibrium system can completely be characterized by 
a small number of state parameters\cite{Wald01LR}. 
The idea of black hole thermodynamics has originated from Bekenstein's 
interpretation\cite{Bekenstein73} 
of Hawking's area theorem\cite{H71} as the 2nd law of thermodynamics, 
as well as from the black hole mechanics due to 
Bardeen, Carter and Hawking\cite{BardeenCarterHawking73}.   
While the latter concerns a stationary equilibrium configuration, 
the former involves a dynamical process concerning 
the total area of all black holes in the universe.  
Therefore, before going into discussion of stationary black holes, 
in this section we shall discuss general circumstances that can 
include some dynamical processes such as a formation and 
evolution of black holes, to which the area theorem becomes relevant.

\subsection{Global structure and area theorem}

First of all, in order to define an isolated black hole in general context,  
one needs to introduce a suitable notion of ``infinity'' 
and associated asymptotic structure. In $4$-dimensions this is usually, 
and elegantly, done in the conformal framework, in which 
an unphysical spacetime $(\tilde M, \tilde g_{ab})$, 
conformally isometric to our physical spacetime $(M,g_{ab})$ 
in $M\cap \tilde M$, plays a role. 
A desired notion of infinity and asymptotic flatness 
can be defined by specifying the behavior of the conformal 
metric ${\tilde g}_{ab}$ near a conformal null boundary 
$\I = \partial \tilde M$. 
If one further imposes an additional condition that every maximally 
extended null geodesic in $M$ has past and future endpoints on 
null boundary $\I$ in $\tilde M$, then $\I$ is divided into disjoint 
sets of the future and past null infinity, $\I^+$ and $\I^-$.  
Such a spacetime is called {\sl asymptotically simple}. 
A spacetime $(M,g_{ab})$ is said to be {\sl weakly asymptotically simple
at null infinity} if $({\tilde M},{\tilde g}_{ab})$ has a neighborhood 
of $\I$ which is isometric to a neighborhood of $\I$ for some asymptotically 
simple spacetime. 
The notion of {\sl strong asymptotic predictability} is then 
defined such that the closure of $M\cap J^-({\I}^+)$ is contained 
in a globally hyperbolic open subset of $\tilde M$. 
A black hole region ${\cal B}$ is defined as the complement of $J^-({\I}^+)$ 
and the future event horizon $H$ as the boundary of ${\cal B}$ in $M$. 
As such, $H$ is a null hypersurface ruled by null geodesics. 
Since ${\cal B}$ is a future set, every null geodesic generator of $H$ is 
future inextendible, but in general admits a past end point. 
With the set of these definitions, general properties of $4$-dimensional 
black holes are studied by using the global method, which consists 
of a number of general results concerning causal structure, behavior 
of causal geodesic congruence, etc, as in Refs.~\citen{HE,Wald84}.

%
A key equation for the global method is the Raychaudhuri equation 
for causal geodesics, which together with certain energy 
condition, governs the occurrence of (a pair of) conjugate points.    
The structure of the Raychaudhuri equation is unchanged in higher dimensions 
as far as (higher dimensional version of) general relativity is considered; 
it takes the form for, e.g., a surface orthogonal null geodesic 
congruence in $D$-dimensions, 
\begin{equation}
 \frac{d }{d \lambda} \theta 
  = - \frac{1}{D-2}\theta^2  - \hat{\sigma}_{ab}\hat{\sigma}^{ab} 
                             - R_{ab}k^ak^b \,,    
\label{eq:Raychaudhuri}
\end{equation}
with $k^a=(d /d \lambda)^a$, $\theta$, $\hat{\sigma}_{ab}$ 
being the tangent of a null geodesic with affine parameter 
$\lambda$ of the congruence, its expansion and the shear. 
Therefore, once well-defined notions of asymptotic flatness 
at null infinity and strong asymptotic predictability 
are formulated in higher dimensions, 
one can apply general results in $4$-dimensions--more specifically, 
Propositions and Theorems in Section 12.2 of Wald\cite{Wald84}--to 
higher dimensions. 
Note that the predictability is needed, in particular to show the area 
theorem without demanding that null geodesic generators of the event horizon 
be complete. 
Note also that in order for a weakly asymptotically simple spacetime in 
$D$-dimensions to be consistent with the asymptotic simplicity 
under the additional condition on maximally extended null geodesics 
mentioned above, each component of $\I$ has to be 
topologically ${\mr}\times S^{D-2}$. This, combined with the topological 
censorship\cite{fsw}, ensures that the domain of outer communication is simply 
connected.

For $D=\mbox{\sl even}$-spacetime dimensions, there exists a stable notion of 
conformal null infinity and weak asymptotic simplicity 
[see Ref.~\citen{HollandsIshibashi2003} for definition], 
and therefore there is no obstruction to apply general results 
[i.e., those in Section 12.2 of Ref.~\citen{Wald84}] 
to higher-even-dimensional spacetimes.

%
However, when spacetime dimension is odd, one needs to be more careful;  
The conformal method for defining null infinity would not in general 
work since the unphysical metric fails to be smooth at conformal 
null infinity $\I$ when radiation is present 
around $\I${}\cite{HollandsWald04}. 
This is essentially due to the fact that the leading fall-off behavior 
of gravitational radiation near null infinity is in proportion to 
a half-integer power of the conformal factor, $\Omega$, when $D$ is odd. 
Therefore, for the case of odd spacetime dimensions, 
we need to formulate a sensible definition of null infinity 
that can be used to define asymptotic flatness and some equivalent notion of 
the strong asymptotic predictability [see Ref.~\citen{TTS10a} 
for such an attempt to define asymptotic flatness in $5$-dimensions 
without using the conformal method]\footnote{ 
One may also want to consider non-asymptotically flat spacetimes, 
such as asymptotically Kaluza-Klein spacetimes.  
For that case, we would not be able to use the standard conformal 
approach to defining null infinity since the compactified dimensions 
shrink to a single point by the conformal transformation. 
We again need to formulate a suitable definition of 
infinity, presumably by dealing with the physical spacetime metric and its 
asymptotic expansion. Note however the stationary case discussed below. 
}. 

In the following when we discuss a black hole in a dynamical, 
non-stationary spacetime, we simply assume that a sensible definition 
of infinity and asymptotic flatness, equivalent to $\I$ 
and weak asymptotic simplicity above, are formulated, and 
we just use the same symbol $\I$ to denote thus defined infinity, 
even if the spacetime dimension is odd and gravitational radiation 
is present near infinity. 
This should be kept in mind when, e.g., a topology changing process 
of cross-sections of the event horizon is considered, 
since such a phenomenon can only occur in non-stationary, 
dynamical spacetime where gravitational radiation is likely to generate.

With the above caveat concerning definitions of null infinity and the 
predictability, 
the standard proof of the theorem [Prop.~9.2.8 in Ref.~\citen{HE}, 
Prop.~12.2.2 in Ref.~\citen{Wald84}] that under the null convergence 
condition, an apparent horizon (see below) 
is contained in the black hole region is straightforwardly generalized 
to arbitrary dimensions.  
In particular, the standard proof of the Hawking's black hole area 
theorem \cite{HE} [Theorem 12.2.6 in Ref.~\citen{Wald84}] is 
generalized to arbitrary higher dimensions.

\subsection{Apparent horizon} 
\label{subsec:AH}
For some purposes, instead of dealing with the event horizon, one is 
more interested in an {\sl apparent horizon} which, in a sense, 
defines a black hole region in a local manner and 
plays a role in particular in numerical studies [See Chapter~9]. 
This notion can be straightforwardly generalized to higher dimensional 
case as discussed below.

Let us consider a partial Cauchy surface, $S$, in $D$-dimensional spacetime 
$M$, which is assumed to be an $(D-1)$-dimensional connected hypersurface 
smoothly embedded in $M$. A closed connected $(D-2)$-surface, $T$, smoothly 
embedded in $S$ is called a {\sl trapped surface} if the expansion of 
the congruence of outgoing light rays orthogonal to $T$ is non-positive 
at each point of $T$. This definition of the trapped surface is well-defined 
only when the notion of ``outgoing" has a definite meaning.
The asymptotic flatness and the orientability of $S$ and $T$ do not give 
the unique definition of `out direction' on $T$. In many cases, it is assumed 
that $S$ is also simply connected. In this case, the out direction of $T$ 
can be defined in terms of the $\boldsymbol{Z}_2$-intersection numbers of 
curves from $T$ to the spatial infinity. 
However, since the simple connectedness of $S$ might be too restrictive, 
it is worth giving another example of a condition that determines the out 
direction of $T$ without an ambiguity. 
Possible such conditions are that $S$ is orientable and that $T$ separates 
$S$ into two disconnected parts. In other words, it is required that 
$S\backslash T=S_{\rm in}\cup S_{\rm out}$ and 
$S_{\rm in}\cap S_{\rm out}=\emptyset$ hold, where $S_{\rm out}$ is 
determined by the property that it contains a neighborhood of the spatial 
infinity. This clearly ensures that $T$ is orientable and defines the out 
direction of $T$ in an obvious way. This condition also applies even 
when $S$ is not simply connected.

%
Under the above conditions, the inner region $S_{\rm in}$ with respect 
to a trapped surface $T$ is called the {\sl inside region $S_{\rm in}(T)$} 
of $T$. 
The inside region of $T$ will be a closed subregion of $S$, whose topological 
boundary in $S$ consists of $T$. Then, the {\sl trapped region of $S$} is 
defined to be the topological sum of $S_{\rm in}(T)$ over all possible $T$. 
The trapped region might not be a closed region or a smooth region of $S$. 
For simplicity, we however only consider the case where the trapped region 
is a smooth closed region of $S$. 
Then, the apparent horizon on $S$ is defined to be the topological boundary 
of the trapped region. It turns out that the apparent horizon 
on $S$ is a {\sl marginally outer trapped surface}, 
or in other words, that the expansion of the congruence of the outgoing light 
rays orthogonal to the apparent horizon is zero everywhere on it\cite{HE}.

%
Now let us consider topological aspects of apparent horizons. 
Hawking\cite{H72LesHouches} has shown under the dominant energy 
condition that the apparent horizon in $4$-dimensional spacetime must be 
diffeomorphic to a $2$-sphere or possibly to a $2$-torus. 
%
%
Hawking's proof of the horizon topology theorem takes two steps: 
First (i) it is derived 
under the dominant energy condition that 
\begin{eqnarray}
\oint_{\Sigma} \mathscr{R} dS \geqslant 
\oint_{\Sigma} 8\pi G T_{ab}l^a (l^b+n^b) \,,
\nonumber 
\end{eqnarray} 
where $\Sigma$ denotes the apparent horizon, $\mathscr{R}$ is the scalar 
curvature with respect to the Riemannian metric induced on $\Sigma$,
$T_{ab}$ is the stress-energy tensor, $l^a$ and $n^a$ respectively are 
the outgoing and the incoming future pointing null vector field orthogonal 
to $\Sigma$. The dominant energy condition requires that 
$T_{ab}l^a(l^b+n^b)\geqslant 0$ holds everywhere. Therefore it follows 
\begin{eqnarray}
\oint_{\Sigma}\mathscr{R} \geqslant 0 \,. 
\label{Yamabe}
\end{eqnarray}
Then (ii) it is appealed to the Gauss-Bonnet theorem, which says that 
the left hand side of the above inequality is $2\pi\chi(\Sigma)$, 
where $\chi(\Sigma)$ is the Euler characteristic number of $\Sigma$. 
It immediately follows that $\Sigma$ must be topologically 
$2$-sphere or $2$-torus. 
The case where the equality in (\ref{Yamabe}) holds, 
if exists, seems to describe an unstable configuration. 
For example, it fails if we add arbitrary small positive 
cosmological constant. Therefore, the toric apparent horizon seems 
implausible, though we need further technical assumptions to exclude it. 
[See Ref.~\citen{Galloway08}, for the issue.]   
In any case, for simplicity, we just assume that the strict inequality 
of $(\ref{Yamabe})$ holds in the following argument.

%
%
Although the inequality (\ref{Yamabe}) in Step (i) holds true also in general 
spacetime dimensions $D \geqslant 4$ without modifications other than 
that $\Sigma$ is $(D-2)$-dimensional as shown in Ref.~\citen{HOY06}, 
the Gauss-Bonnet theorem in Step (ii) can apply only in $4$-dimensions.   
Noting this fact, Galloway and Schoen\cite{GallowaySchoen06,Galloway08} 
have shown, as a natural generalization of Hawking's topology result mentioned 
above to higher dimensions, that an apparent horizon must admit 
a Riemannian metric of positive scalar curvature. 
This in general restricts possible topological 
or differentiable structure of the apparent horizons. 
Now the problem has reduced to a purely geometric argument, 
and we can apply the mathematical results in the standard differential 
geometry. 
For example, in $5$-dimensions, $\Sigma$ must be either 
$3$-sphere with possibly identifications, or $S^2\times S^1$, 
or a finite connected sum of them. 
Of course, an apparent horizon in $5$-dimensional spacetime must be 
homeomorphic to one of them.

Thus, we have many possibilities for the topology of the apparent horizon. 
Since it is known that a spatial section of the event horizon 
in a stationary spacetime coincides with the apparent horizon,  
this result indicates that there might be a rich variety for 
the final equilibrium configuration of black holes in higher dimensions. 
In fact, such an example has been explicitly constructed by Emparan 
and Reall\cite{ER02b}. 
They have found a black hole solution whose black hole horizon is 
diffeomorphic with $S^2\times S^1$ in $5$-dimensional Ricci flat spacetimes.

It is interesting to note that by applying a similar argument used in 
the derivation of \eqref{Yamabe} to asymptotically (locally) 
anti-de Sitter black holes, one can obtain a topology dependent lower bound 
for the area, hence entropy, of the black hole\cite{Gibbons99,Woolgar99} 
[see also Refs.~\citen{CaiGalloway01,GallowayOMurchadha08,Racz08}].

\subsection{Topological structure of dynamical black hole horizons}
Next, let us briefly see the dynamical aspects of black hole horizons. 
We focus on the time evolution of the topological structure of event horizons.
The future event horizon $H$ is 
generated by null geodesics without a future end point.   
The set consisting of all the points where two or more null geodesic 
generators of $H$ intersects is called the {\sl crease set} $F$ of $H$. 
Each null geodesic generator of $H$ has a past end point on the closure 
$\overline F$ of $F$, or else it does not have a past end point. 
We here consider black holes which is created within finite past, 
or in other words, we only consider the case where any geodesic generator 
has a past end point on $\overline F$ in the following. 
The event horizon $H$ may consists of several connected components, which 
might occur, for example, when the final state is represented by one of 
the Majumdar-Papapetrou solutions to the Einstein-Maxwell equations 
describing equilibrium states of several extremely charged black holes. 
However, it is enough to consider the case where $H$ consists of one 
connected component, as we assume here, for generalization to the case with 
many black holes is straightforward. 
From the definition of $F$, it immediately follows that the crease set $F$ 
is an acausal set in $H$\cite{Siino98}, 
that is any two points in $F$ are causally separated.

More stringently, the following theorem holds in $D$-dimensional 
spacetimes: if the partial Cauchy surface is $\boldsymbol{R}^{D-1}$ and 
the spatial section of $H$ is homeomorphic with the $(D-2)$-sphere 
at sufficiently late times, the crease set $F$ is a contractible 
space. 
In particular, it immediately follows that $F$ is arcwise connected. 
This gives a simple picture for the time evolution of event horizons. 
In fact, it is easy to show that there is a time slicing of $M$ in which 
a black hole is created at a point, and the spatial section of $H$ 
is homeomorphic with $(D-2)$-sphere in subsequent times.

Another interesting viewpoint of Siino\cite{Siino98} is that
the number of black holes is a gauge dependent notion.
The event horizon is clearly a gauge invariant object,
defined only by the terminology in the causal structure of spacetimes. 
However, whenever we count the number of black holes, we first prepare 
appropriate  time slicing of $M$ to do it, which is clearly a gauge dependent 
procedure. For example, let us consider the head on collision of two spherical 
black holes into one. This situation will realize in an axisymmetric 
configuration, where the axis of symmetry is the straight line along which 
two black holes take trajectories. Then, we might expect that the spatial 
section of $H$ before the collision consists of two connected components. 
In fact, it is not always the case. In this situation, the crease set $F$ 
will be a topological line segment in the world surface of the symmetric 
axis. Since $F$ is an acausal set, it is just a spacelike line segment. 
Now, we can take a partial Cauchy surface $S$, such that entire $F$ is 
included in $S$. 
Then, if we consider a time slicing of $M$ including $S$ as a time slice, 
we will not see disconnected black holes at any time. In other words, 
this spacetime can be interpreted as just a creation of a black hole. 
Furthermore, since $F$ is a spacelike line segment as noted above, 
we can also choose a partial Cauchy surface $S'$, where $S'$ intersects 
$F$ transversally at many points. Then, we will conclude that there are 
more than two black holes on $S'$.

A better theorem has been obtained by Ida\cite{Ida10}. The event horizon $H$ 
is an $(D-1)$-dimensional topological submanifold of $M$, which follows from 
the fact that it is the boundary of a causal past $\dot J^-(\mathscr{I}^+)$ 
in $M$\cite{HE}. Note that it may not be smoothly embedded in $M$, 
but it has a wedge-like structure at $\overline F$ in general. 
The following theorem gives a relationship between the topological structure 
of $F$ and that of the topological manifold $H$: 
The crease set $F$ in $H$ is a deformation retract of $H$. 
Then, it immediately follows that the homotopy type of $F$ and $H$ coincides. 
In particular, the crease set must be arcwise connected, which has already 
been noted by Siino\cite{Siino98}.  
However, this theorem also applies to homotopy types of higher order. 
For example, $F$ must be simply connected if $H$ is.

It also holds\cite{Ida10} that the final topology of the spatial section of 
the event horizon is diffeomorphic with the $\epsilon$ neighborhood of $F$ 
in $H$, where $\epsilon$ neighborhood is defined by introducing a Riemannian 
metric on $H$ determined in terms of the differentiable structure on $M$ 
(of course, it is not the induced metric). 
In this sense, the topology of the black hole is determined 
by the crease set $F$.

Next, let us discuss several forbidden processes in topology change of 
black holes. 
A well known is the no bifurcation theorem of black hole due to 
Hawking\cite{HE}, which says that a black hole does not evolve into
several black holes by bifurcating.
More precisely, it states that
$J^+(\mathscr{B}_0)\cap S_1$ is arcwise connected, where 
$\mathscr{B}_0$ denotes a connected black hole region on
a partial Cauchy surface $S_0$ 
[so $\mathscr{B}_0 \subseteq {\cal B} \cap S_0$], 
and $S_1$ is a partial Cauchy surface at later
time, that is entirely lying within  $I^+(S_0)$. 
This gives one of the most basic forbidden processes for black holes. 
Since  it seems that there are more possibilities for black hole topology in higher dimensional spacetimes,
one might expect that there might be other forbidden processes than this.
This problem is considered in Ref.~\citen{IdaSiino07} in the language of the Morse theory.
First, note that the topology of a black hole does not change at sufficiently late times, where
time slices do not intersect the crease set $F$,
since the null geodesic generators of the event horizon naturally defines the diffeomorphism between the black hole horizons on different time slices.
In other words, the black hole horizon can change its topology only at the instant when the time slice intersects the crease set $F$.
From this viewpoint, the topology change of black holes can be regarded as a local process occurring in a small neighborhood of $F$.
 In order to apply the Morse theory to the event horizon, we first introduce a smoothed event horizon $\widetilde H$, which is a differentiable manifold smoothly embedded in $M$, such that this smoothing procedure preserves 
the topological structure of black hole horizon on each time slicing. Furthermore, we consider a situation where the time function $t$ induced 
on the smoothed event horizon $\widetilde H$ gives a Morse function on $\widetilde H$, that
each critical point of $t$, defined as such point where
$\partial_i t=0$ holds, where the partial derivatives are those on $\widetilde H$,
is a regular critical point, where
the ${\rm det}(\partial_i\partial_j t)\ne 0$ holds.  
This technical assumption is not always justified, although it is always possible for any black hole spacetime to find such time slicings that the above operations are justified. 
Once the time function $t$ on $\widetilde H$ is recognized as a Morse function on $\widetilde H$,
the topology change of the black hole horizon can be seen as a local process occurring at a critical point of $t$ on $\widetilde H$. This allows us only to consider the evolution of a local black hole region  around a critical point, which can in fact be regarded as a process in the Minkowski spacetime.
In this way, it turns out that there are many forbidden processes for the topology change of black hole horizons. 

Here, we provide an example among these. 
Let us consider the process where an initial black hole horizon homeomorphic with $S^{D-1}$ 
change its topology into $S^{D-2}\times S^1$. 
There are topologically distinct two possible processes for this to occur at a specific instant of time. One possibility is such that a pair of horns grows, one from the North pole, and the other from the South pole of the sphere, and then merges each other at a point.
The other possibility is that the sphere punctures in such a way that the North pole and the South pole meet each other from the inside of the sphere.
The former process can actually occur, but the later turns out to be a forbidden process.

Finally, we consider the topology of crease set of the event horizon in 5-dimensional spacetimes.
In this case, the topology of the black hole at late times is explicitly known.
Note that if the black hole horizon in final equilibrium state is diffeomorphic with a 3-manifold $N$,
the event horizon $H$ is diffeomorphic with the interior of a 4-manifold bounded by $N$.
Furthermore, $H$ must be embedded in the partial Cauchy surface. They can be seen by noting that $H$ is an acausal set and that a timelike vector field without zero, which always exists in any time orientable Lorentzian manifold, naturally defines a diffeomorphism of $H$ into a portion of a partial Cauchy surface.

First, let the partial Cauchy surface $S$ be $\boldsymbol{R}^4$. Then, $N$ must be $S^3$, $S^2\times S^1$, or a connected sum 
$\#_m (S^2\times S^1)$.
The lens space black hole is not realized in this case, for it cannot be embedded in 
$\boldsymbol{R}^4$.
If the crease set $F$ is a point, then $H$ is an open 4-disk and $N$ is a 3-sphere.
If $F$ is a circle, then $H$ is an open solid torus and $N$ is $S^2\times S^1$.
The crease set for 
$N\simeq \#_m (S^2\times S^1)$ is given by a chain of $m$ circles.
The chain of circles is here defined as follows. Let $\{C_i\}$ ($i=1,2,\cdots,m$) be a sequence of circles. Then the chain of $m$ circles is the space $\bigcup_i C_i$ with identification of a point in $C_i$ and a point in $C_{i+1}$ for each $i=1, \cdots, m-1$.

To realize the lens space black holes, we must consider the partial Cauchy surface that is not $\boldsymbol{R}^4$.
A simplest example is obtained when $H$ is a 2-dimensional real vector bundle over the 2-sphere. The topology of such spaces is classified by the Euler number $p\in\boldsymbol{Z}$,
which corresponds to the Dirac monopole number of the associated principal
$U(1)$ bundle.
Hence we denote it by $E_p$. Such an event horizon $H\simeq E_p$ emerges when the crease set is the zero section of $E_p$, that is the 2-sphere. The vector bundle $E_p$ is clearly bounded by $S^1$ bundle over the 2-sphere, that gives the final topology of the black hole horizon.
The vector bundle $E_0$ is just a product space $S^2\times \boldsymbol{R}^2$, which is bounded by $S^2\times S^1$. Thus the black ring can also emerge from $F\simeq S^2$.
The vector bundle $E_1$ is bounded by the 3-sphere. This is known as the Hopf fibration of $S^3$.
Hence, the spherical black hole can also emerge from $F\simeq S^2$.
The Kaluza-Klein black hole discussed in Ref.~\citen{IshiharaMatsuno06} belongs exactly to this type.
For general $p$, $(|p|\ge 2)$, the boundary of $E_p$ becomes the lens space $L(p,1)$. Thus, the lens space black hole of this special type can emerge 
from $F\simeq S^2$. 
In fact, the black lens spacetime in Ref.~\citen{Ida10} 
corresponds to $H\simeq E_2$. 
General black lens horizon $\simeq L(p,q)$ is modelled by $H$, which is 
a plumbing of a sequence $\{E_{p_i}\}$, namely the chain of vector bundles 
in an appropriate sense. 
This is discussed by Ida\cite{Ida10}, in which it is shown that the 
$5$-dimensional Kastor-Traschen spacetime\cite{Kastor&Traschen93} 
includes black lens spacetime with the event horizon consisting of 
plumbing of $m$ vector bundles $E_2$, to result in $L(m+1,1)$ black lens 
horizon.

\section{Stationary black holes in higher dimensions} 
\label{sec:stationary}

Now we turn to stationary black holes in a smooth, strongly causal 
spacetime $(M,g_{ab})$. By {\sl stationary} we mean that there exists 
a Killing vector field $t^a$ whose orbits are complete everywhere 
in the spacetime and are timelike at least at large distances. 
The completeness is important in the following arguments, 
as otherwise $t^a$ would fail to generate an isometry group, $\phi_t$.  
In this case, there is no problem to define the notion of asymptotic flatness 
at null infinity of the standard topology $\I\approx {\mr } \times S^{D-2}$ 
in both even and odd dimensions within the standard conformal framework.  
For more general cases including asymptotically Kaluza-Klein spacetimes,
one may apply the method using an {\sl asymptotically flat end}, 
$S_{\rm ext}$, of a partial Cauchy surface $S$, on which 
the induced metric and extrinsic curvature satisfy some appropriate fall-off 
conditions. 
[See Refs. \citen{CW94,CC08,Chr09} for details.] 
The one-parameter group of diffeomorphism of $t^a$ 
defines an asymptotically flat exterior region by 
$M_{\rm ext}=\cup_{t\in {\mr}} \phi_t (S_{\rm ext})$. 
The black hole region is then ${\cal B}= M \setminus J^-(M_{\rm ext})$.  
One can choose a cross section $\Sigma$ of $H = \partial {\cal B}$ 
so that $t^a$ is everywhere transverse to $\Sigma$ 
on $H \cap I^+(M_{\rm ext})$. By Lie-dragging $\Sigma$ over 
$H$ along the orbits of $t^a$, one can define 
a foliation, $\phi_t(\Sigma)$. 
Then, using the Raychaudhuri equation, (\ref{eq:Raychaudhuri}), 
the vacuum Einstein equations, and the same arguments as used to prove 
the area theorem, one can show that the expansion and shear of the null 
geodesic generators of the event horizon vanish as asserted 
in Prop.~9.3.1 of Ref.~\citen{HE}. 
It then follows that all cross-sections of the event horizon of 
a stationary black hole are isometric, as far as they are chosen 
in the manner just mentioned above. 
[Compare with the non-stationary case discussed in the previous section.]

%
It may be instructive to start with briefly recapitulating basic steps 
of the uniqueness proof for asymptotically flat, stationary vacuum black holes 
in $4$-dimensions and then to describe attempts to generalize those steps to 
higher dimensions.  
The proof in $4$-dimensions goes roughly as follows: 

\medskip 
\noindent
{\sl (i) Topology theorem: } 
The first step in the proof is to show that each connected component of 
a cross section of the event horizon of an asymptotically flat, 
stationary black hole must have spherical topology. 
This was first shown by Hawking\cite{H72} 
[see also Prop.~9.3.2 of Ref.~\citen{HE}, and the previous argument 
in subsection~\ref{subsec:AH}], in which a variational 
argument, the dominant energy condition, and the Gauss-Bonnet theorem 
are used. 
A stronger proof was later given by Chru\'sciel and Wald\cite{CW94b} 
by using the topological censorship\cite{fsw}, the null energy condition, 
and cobordism theory.  
The topological censorship, which requires the weak energy 
(or null convergent) condition, is used to show the simple connectedness 
of the domain of outer communications. Then basic result of cobordism theory 
yields that a simply connected Riemannian $3$-manifold with a boundary 
$S^2$ at one end (at infinity) must have $S^2$ at another end (at horizon).

\medskip 
\noindent
{\sl (ii) Rigidity theorem: } 
The next step is to show that a stationary black hole is either static or 
axisymmetric. 
One can show, by using the result of topology theorem~{\sl (i)} that 
(1) there exists a Killing field $K^a$ which is normal to $H$, 
hence $K^aK_a =0$ on $H$, and furthermore that (2) if the black hole 
is rotating, there must exist another 
independent Killing vector field $\phi^a$, 
which commute with $K^a$ and generates an isometry group $U(1)$. 
Thus, the spacetime isometry group is at least ${\mr}\times U(1)$. 
[See Prop.~9.3.6 of Ref.~\citen{HE}. See also Refs.~\citen{Carter72,Chr97}.] 
(1) establishes that the event horizon is a {\sl Killing horizon}. 
An immediate consequence is the constancy over $H$ of the surface 
gravity, $\kappa$, defined by $K^b\nabla_bK^a = \kappa K^a $ on $H$. 
Since $\kappa$ defines the Hawking temperature, this establishes 
the basis of the $0$th-law of black hole thermodynamics.  
(2) establishes that the spacetime, when rotating, is {\sl axisymmetric}. 
Since it implies that the generators of the event horizon are 
{\sl rigidly} rotating with respect to infinity, it is called 
the rigidity theorem.

When $t^a$ is null on $H$, i.e., non-rotating, we appeal to the staticity 
theorem of Ref.~\citen{SW92} to show that the spacetime is static. 
Then, we apply the theorem of Israel\cite{Israel67wq} to show 
the uniqueness of such a solution. We will discuss this case 
and its higher dimensional generalizations in detail 
in Section~\ref{sec:unique:static}.

\medskip  
\noindent
{\sl (iii) Non-linear sigma model and global divergence identity: }
The next step is, by using the two isometries obtained in Step {\sl (ii)}, 
we reduce the Einstein equations to a certain type of non-linear sigma 
model on a $2$-dimensional base space. We see that the target space 
of the sigma-model is homogeneous and non-compact, thus having 
a non-positive sectional curvature.  
In fact it is $SL(2,\mr)/SO(2)\cong {\boldsymbol H}^2 $ for the vacuum case. 
The reduction is achieved with the help of the Weyl-Papapetrou coordinates.  
We also construct a (Mazur's) divergence identity [see also 
Bunting's in section~\ref{sec:unique:stationary}  
for non-homogeneous case], expressed in terms of a coset matrix. 
Note that this is a global identity as the target space is homogeneous.

\medskip 
\noindent
{\sl (iv) Boundary value analysis: } 
We show by using the global divergence identity that 
given boundary conditions that correspond to specifying 
a set of asymptotic conserved charges, the solutions are uniquely determined. 

\medskip 
Technical details of the uniqueness proof are found in 
literature~\citen{CC08,Heusler96}. 
For the rest of this section we shall review how the first two steps,     
{\sl (i)} and {\sl (ii)}, are generalized to higher dimensions. 
Higher dimensional generalizations of the last two steps, {\sl (iii)} and 
{\sl (iv)}, will be discussed in more detail in 
Section~\ref{sec:unique:stationary}.

%
\subsection{Topology theorems in higher dimensions} 

\subsubsection{Topology of cross sections of the event horizon}
\label{subsec:topology}

As mentioned above, all cross-sections $\Sigma$ of 
the event horizon $H$ of a stationary black hole are 
isometric---thus slice independent---in $I^+(M_{\rm ext})$, 
and this property holds true in higher dimensional case as well. 
Furthermore an apparent horizon of a stationary black hole coincides 
with the event horizon. Therefore the topology theorem for an apparent 
horizon discussed in the previous 
section\cite{H72LesHouches,GallowaySchoen06,Galloway08}  
applies to the even horizon:

\medskip 
\noindent
{\bf Theorem 3.1:}~[Reference \citen{GallowaySchoen06,Galloway08}]  
{\em Cross sections of the event horizon in stationary black hole spacetimes 
obeying the dominant energy condition are of positive Yamabe type.} 

\medskip 
This implies that cross sections $\Sigma$ must be able to carry 
a metric of positive scalar curvature. In particular, the case of 
toroidal cross sections has been excluded\cite{Galloway08}. 
Accordingly, in $5$-dimensions, $\Sigma$ must be a connected sum of $S^3$, 
$S^2\times S^1$, and some quotient space of $S^3$, including a Lens space. 

A higher dimensional generalization of the proof of Chru\'sciel 
and Wald\cite{CW94b}, in which the topological censorship and cobordism 
is used, has also been discussed by Helfgott, Oz and Yanay\cite{HOY06}. 
However, the topological constraint thus obtained is less restrictive 
than that of Refs.~\citen{GallowaySchoen06,Galloway08} above.

Further topological restrictions can be obtained when a stationary 
black hole admits extra symmetries. 
Suppose a $D$-dimensional stationary black hole spacetime admits 
$(D-3)$-rotational isometries. 
(Note that for $D \geqslant 6$, this assumption is no longer compatible 
with the standard notion of asymptotic flatness with $S^{D-2}$ 
spatial infinity. It is, instead, relevant to asymptotically 
Kaluza-Klein spacetimes.) 
It is known that such a stationary, multi-axisymmetric spacetime 
possesses a set of invariants---called the {\sl rod-structure} of 
Harmark\cite{Harmark04,Harmark-Olesen05}---which consists of a set of 
positive numbers and vectors living in a boundary segment of 
the $2$-dimensional factor space of $M$ by the isometry and defines 
the structure of the `zero sets' of the Killing vector fields.  
The rod-structure encodes, especially, information about the horizon topology. 
Given a rod on the boundary segment, the associated 
positive number corresponds to the invariant length of the rod,  
and the vector associated with the rod (apart from that corresponding  
to the horizon) specifies which linear combination of the (rotational) 
Killing vector fields vanishes in the rod. 
[See Chapter~4 for details of the rod-structure. See also Ref.~\citen{ER02a} 
for earlier work on the rod-structure considered in static spacetimes]. 
A refined version of the rod-structure---called 
the {\sl interval structure}---has been proposed in 
Refs.~\citen{HollandsYazadjiev08,HollandsYazadjiev08b}. 
In particular, it has been made clear that the determinant of the matrix 
made of two integer valued vectors associated with adjacent rods 
characterizes the horizon topology. 
For example, it has been shown that for a $5$-dimensional asymptotically flat, 
stationary vacuum black hole with two axial symmetries, 
each connected component of the horizon cross section must be topologically 
either $S^3$, $S^2\times S^1$, or a Lens-space. 
The concrete relation between the possible horizon topology and 
the interval structure 
has been given [see Prop.~2 of Ref.~\citen{HollandsYazadjiev08}]. 
This result has been generalized to a more general case of stationary 
asymptotically Kaluza-Klein black holes with $(D-3)$ rotational 
isometries\cite{HollandsYazadjiev08b}.

The rigidity theorem states that any stationary rotating black hole 
spacetime must be axisymmetric. This holds true in higher dimensions 
as well [see next subsection] but it guarantees the existence of 
only a single $U(1)$ isometry, irrespective of the number of spacetime 
dimensions. 
Therefore there could exist a stationary black hole solution in higher 
dimensions that has precisely a single $U(1)$ isometry, as conjectured 
by Reall\cite{Reall03}. 
Having only a single $U(1)$ isometry, we would have less restrictive 
constraint on the horizon topology than the case of having more than 
one axial symmetries, say $U(1)^{D-3}$ ($D \geqslant 5$) discussed above, 
but more restrictive constraint than what is implied by merely knowing 
that $\Sigma$ is of positive Yamabe type\cite{GallowaySchoen06,Galloway08}. 
It has been shown in Ref.~\citen{HHI10} that the horizon cross section of 
$5$-dimensional asymptotically flat, stationary vacuum black holes can be 
either connected sum of Lens spaces and handles ($S^1\times S^2$), 
or the quotient of $S^3$ 
by certain finite subgroups of isometries with no handles. 
The latter horizon manifold includes Prism manifolds, quotients of 
the Poincare homology sphere, and a various Seifert fibred spaces 
over $S^2$ [see Table~1 of Ref.~\citen{HHI10} for the list of 
possible horizon topologies].

The notion of the rod-structure---which requires $(D-2)$ mutually 
commuting isometries---has been generalized to 
the {\sl domain structure}\cite{Harmark09}, which applies to the case 
of any spacetime dimensions and any numbers of Killing vector fields. 
Thus, in particular, the domain structure can be used to characterize 
a stationary black hole geometry with fewer rotational symmetries 
than $(D-3)$ rotational symmetries.

\subsubsection{Topology of the domain of outer communications}
Compared with the horizon topology, perhaps less appreciated is 
the topology of the domain of outer communications.
The topological censorship\cite{fsw,CGS09,Galloway01} 
states that 
any curve in the domain of outer communication with endpoints 
in the asymptotic region $M_{\rm ext}$ can be deformed to a curve entirely 
within $M_{\rm ext}$. 
Therefore, for a black hole in an asymptotically flat spacetime 
in the standard sense, the domain of outer communication 
must be simply connected. 
However, in higher dimensions, the simple connectedness itself does not 
completely determine the topology of a partial Cauchy surface $S$ 
in the domain of outer communication, though it is likely to be 
$\mr^{D-1} $ minus a compact manifold ${\mathscr B}$ that describes 
the black hole region on $S$.

%
In $5$-dimensions, it has been shown in Ref.~\citen{HollandsYazadjiev08b} 
by using the results of Ref.~\citen{orlik1} that 
when a stationary, rotating black hole spacetime admits two axial 
isometries $U(1)\times U(1)$, the domain of outer communication 
has topology $\mr \times S$, where for some $n,n'\in\mn$, 
\ben
 S \cong \Bigg( \mr^4 \, \# \, n \cdot (S^2 \times S^2) \, \# \, 
n' \cdot(\pm \boldsymbol{C} P^2) \Bigg) \setminus \mathscr{B} \,, 
\een
where $\mathscr{B}$ denotes a compact manifold with boundary 
$\partial {\mathscr B}= \Sigma$, an intersection 
of a partial Cauchy surface $S$ and the black hole region ${\cal B}$.   
It has also been shown in Ref.~\citen{HollandsYazadjiev08b} that 
as in the case of the horizon topology, the topology of domain of outer 
communication can be completely specified in terms of the interval structure. 
Note also that if the black hole is non-rotating, then it is static 
due to the result of Ref.~\citen{SW92}. Then, the solution is shown 
to be isometric to the $5$-dimensional Schwarzschild 
spacetime\cite{Hwang98,GIS2,Gibbons02b}. 
[See Section~\ref{sec:unique:static}.] 
Therefore $S \cong \mr^4 \setminus \mathscr{B}$.

For $5$-dimensional stationary black holes with only one $U(1)$ isometry, 
in accord with the rigidity theorem, 
one might expect to have less restrictive constraints on the possible 
topology, as in the even horizon case. 
However, it is shown by Ref.~\citen{HHI10} that 
the topological constraint on domain of outer communication of this 
less symmetric case is essentially the same as the case with two rotational 
isometries ${ U(1)\times U(1)}$.

\subsection{Rigidity theorem in higher dimensions} 
\label{subsec:rigidity}
A higher dimensional generalization of the rigidity theorem has been made 
by Hollands, Wald and one of the present authors\cite{HIW07}, as well as 
by Moncrief and Isenberg\cite{MI08}. In this subsection we shall review 
the results obtained in Ref.~\citen{HIW07}. 

Hawking's proof of the rigidity theorem relies in an essential way 
upon the fact 
that the spacetime dimensions is $4$, and for this reason, 
it is a priori not at all obvious whether the rigidity theorem can 
be generalized to higher dimensional case.  
Let us start with a brief sketch of the rigidity proof 
and see how the spacetime dimensionality enters the proof. 
Since the stationary Killing vector field $t^a$ generates a one-parameter 
group of isometries, it must be tangent to the event horizon, $H$. 
When $t^a$ is not normal to $H$ (i.e., not tangent to the null 
geodesic generators of $H$), the black hole is said to be {\sl rotating}. 
In this case our task is first to show that there exists an additional 
Killing vector field, $K^a$, that is normal to $H$ on $H$, 
besides $t^a$, and next, having the desired $K^a$, 
to find (a linear combination of) axisymmetric Killing fields
$\varphi_{(i)}^a$ by $t^a = K^a+ \Omega_{(i)}\varphi_{(i)}^a $, with $\Omega_{(i)}$ 
being some constants. Such a desired Killing vector field, $K^a$, 
is (1) first to be constructed locally in a neighborhood of $H$, and 
(2) then to be extended to the domain of outer communication. 
For the purpose of Step (2), we assume that the spacetime metric and 
matter fields be real analytic and then show 
that the `Taylor expansion' at $H$ satisfies   
\ben
 \underbrace{\pounds_\ell \, \pounds_\ell \, \cdots \,
 \pounds_\ell}_{m \,\, {\rm times}} 
(\pounds_K \Phi) = 0 \,, \quad m=0,1,2, \dots  
\quad \mbox{on $H$} \,,  
\label{coeff:Taylor}
\een 
where $\Phi$ collectively denotes all relevant physical fields 
and $\ell$ denotes some vector field that is transverse to $H$. 
Then, by analytic continuation we extend $K^a$ to the entire spacetime.

\begin{figure}
   \centerline{\includegraphics[width=4.5cm,height= 3.3cm]{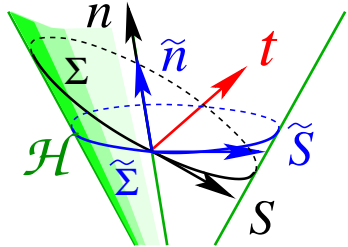}}
   \caption{\small 
            The decomposition $t^a=n^a+s^a$ depends on the choice of 
            foliation $\Sigma$. We wish to find a `correct' foliation 
            $\tilde \Sigma$ so that the corresponding ${\tilde n}^a =K^a$ 
            satisfies---as a candidate Killing field---the desired 
            properties~(i)--(iii). 
            } 
   \label{fig:C}
\end{figure}

\begin{figure}
   \centerline{\includegraphics[width=2.5 cm,height= 3cm]{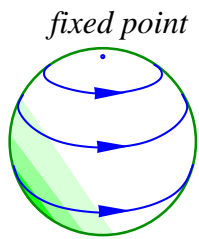} }
   \caption{\small A horizon cross-section $\Sigma$ for a $4$-dimensional 
            black hole. The blue lines denote 
            the flow of $s^a$, which has a fixed point 
            and closed orbits on $\Sigma \approx S^2$.}
   \label{fig:2Sphere}
\end{figure}

In the following we focus on Step (1); we find a candidate $K^a$ on the event 
horizon. The properties that the desired, candidate Killing field $K^a$ 
should possess on $H$ are: 
(i) $\pounds_t K^a=0$ and $K^aK_a =0$, 
(ii) $\pounds_K \Phi=0$, and (iii) $K^c\nabla_c K^a = \alpha K^a$ 
with $\alpha$ being constant, which is to be identified with the surface 
gravity $\kappa$. 
Now let us choose a foliation of $H$ by compact cross-sections 
$\Sigma$ and decompose the stationary Killing vector $t^a$ 
on $H$ with respect to $\Sigma$ as $t^a= n^a + s^a$, 
where $n^a$ is null and $s^a$ spacelike, tangent to $\Sigma$. 
(One can construct a well-behaved foliation by first choosing a single 
cross-section $\Sigma_0$, and then Lie-dragging $\Sigma_0$ over $H$ 
by the isometry of $t^a$.)
Then, it is straightforward to check that $n^a$ satisfies (i) and (ii). 
However, there is a prior no reason that $\alpha$ with respect to $n^a$ 
needs be constant, since the decomposition $n^a=t^a-s^a$ depends upon 
the choice of $\Sigma$. (See figure~\ref{fig:C}.) 
Therefore our task is, starting from an arbitrarily chosen $\Sigma$, 
to find the `desired' foliation $\tilde {\Sigma}$ that gives rise to 
${\tilde n}^a=K^a$ with $\tilde \alpha$ being constant over $H$, 
satisfying the property~(iii). 
It turns out that to find such a desired foliation $\tilde \Sigma$, 
one has to integrate along the orbit of $s^a$ 
a set of two ordinary differential equations on $\Sigma$, both of which 
can be written in the form 
\ben
  \pounds_s \Psi = J \,, 
\label{eq:*}
\een
where $\Psi$ corresponds either to a function that 
correctly `normalizes' $n^a$ to obtain ${\tilde n}^a=K^a$ or to a coordinate 
function that defines the desired foliation, $\tilde \Sigma$, 
and where $J$ is some smooth function on $\Sigma$. [See {Lemma~2} of
Ref.~\citen{HIW07}.]
Now when solving this equation with respect to $\Psi$, 
the spacetime dimension and the topology of $\Sigma$ play a crucial role. 
For $4$-dimensions, the cross-section $\Sigma$ must be topologically 
$2$-sphere due to the topology theorem. It then immediately follows 
that the flow of $s^a$ on $\Sigma$ must have a fixed point, $p$, 
as the Euler characteristic of $\Sigma \approx S^2$ is non-zero.  
[See figure~\ref{fig:2Sphere}].
\begin{figure}
   \centerline{\includegraphics[width=4 cm,height= 3cm]{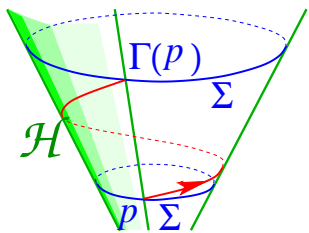} }
   \caption{\small 
            An event horizon $H$ of a $4$-dimensional stationary 
            black hole and a foliation of $H$ by $\Sigma$. 
            The red line denotes an orbit of $t^a$ on $H$. 
            A point $p$ on a null geodesic generator of $H$ is mapped by 
            a discrete isometry $\Gamma_t$ to a point $\Gamma_t(p)$ 
            on the same null geodesic generator.}  
   \label{fig:Gamma}
\end{figure}
Now the (infinitesimal) action of $s^a$ on any $2$-vector, 
$v^a$, on the tangent space at the fixed point $p$ (where 
$s^bD_bv^a =0$), is $\pounds_s v^a = - (D_bs^a) v^b$. 
Since $D_b s^a$ is a $(2\times 2)$ anti-symmetric matrix, 
the action of $s^a$ describes an infinitesimal `rotation' 
on the tangent space at $p$. Therefore all the orbits of $s^a$ 
must be closed with a certain period $P$. 
Then, by integrating Eq.~(\ref{eq:*}) along a closed orbit of $s^a$ one can 
always find a well-defined solution $\Psi$ which gives rise to 
our desired foliation $\tilde \Sigma$ and the horizon normal Killing field 
$K^a={\tilde n}^a$.

%
That the orbits of $s^a$ are closed implies that $t^a$ generates 
a discrete isometry, $\Gamma_t$, which maps each null generator of 
the event horizon into itself [see figure~\ref{fig:Gamma}]. 
Therefore if we 
identify points in the spacetime that differ by the action of $\Gamma_t$ 
with period $t=P$, the event horizon becomes a compact null hypersurface 
ruled by closed null geodesic generators. Then, one can invoke 
Isenberg and Moncrief's symmetry theorem for compact Cauchy 
horizons~\cite{MI83,IM85}, which provides the desired additional 
Killing field normal to the horizon as shown in Ref.~\citen{FRW99}.  
The discrete isometry $\Gamma_t$ implies that the surface gravity is 
set to be 
\ben
  \kappa = \frac{1}{P}\int_0^P \alpha[\phi_s(x)] d s \,,   
\een
where $\phi_s(\cdot)$ denotes the flow generated by $s^a$ 
and $x \in \Sigma$.

%
In higher dimensions $D>4$, however, cross-sections $\Sigma$ of 
the event horizon can admit non-trivial topology, and there is no reason 
that the isometries of $\Sigma$ generated by $s^a$ need have closed 
orbits even if $s^a$ vanishes at some point $p \in \Sigma$. 
(This would be the case even in $4$-dimensions if the horizon 
topology were non-spherical, e.g., torus). 
An example is supplied by considering a $5$-dimensional 
Myers-Perry black hole solution~\cite{MP86}, 
whose event horizon cross-section is topologically $\Sigma \approx S^3$. 
The solution admits two rotational Killing fields, 
$\varphi^a_{1}$, $\varphi^a_{2}$ and their linear combination 
provides $s^a$ on $\Sigma$. 
If we choose two rotational parameters in the linear combination 
so that their ratio becomes incommensurable, then the orbits of 
$s^a$ do not have a closed orbit on $\Sigma$. 
Therefore, in general, there is no guarantee that one can find 
a well-defined solution, $\Psi$, for the higher dimensional black hole case.  
%
\begin{figure}
   \centerline{\includegraphics[width=3.7 cm,height= 2.8cm]{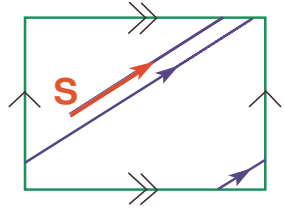}}
   \caption{\small 
            The fundamental region of a $2$-dimensional torus. Two-dimensional 
flat torus admits two Killing fields, each of 
which has closed orbits, but their linear combination $s^a$ does not 
necessarily have closed orbits. 
            } 
   \label{fig:torus-id}
\end{figure}
%
This problem may be illustrated by a simple, lower-dimensional case; 
let us consider the case in which $\Sigma$ is $2$-dimensional flat torus 
and attempt to solve the same type of equation,~(\ref{eq:*}), on $\Sigma$ 
along a non-closed orbit of $s^a$ on $\Sigma$. The torus $\Sigma$ has 
two Killing fields $\tau^a_1 = (\partial /\partial \tau_1)^a$ and 
$\tau^a_2= (\partial/\partial \tau_2)^a$, each of which has closed 
orbits on $\Sigma$. Then, $s^a$ with non-closed orbits can be expressed 
as a linear combination 
\ben
 s^a = \Omega_1 \tau_1^a  - \Omega_2 \tau_2^a \,, 
\een
with the ratio, $\Omega_1/\Omega_2$, of the two coefficients being 
an irrational number. 
Then, it immediately follows that in terms of Fourier transform 
$\widehat{J} ({\boldsymbol x},m_1,m_2)$ of $J({\boldsymbol x}, \tau_1,\tau_2)$, 
a {\sl formal} solution $\Psi$ to \eqref{eq:*} is given by 
\ben
\Psi({\boldsymbol x}) 
  = i\!\sum_{m_1, m_2} \!
    \frac{\widehat{J}({\boldsymbol x},m_1,m_2)}{m_1\Omega_1 - m_2 \Omega_2} 
  = i\!\sum_{m_1, m_2} \!
     \frac{ \widehat{J}({\boldsymbol x},m_1,m_2) 
          }{m_1\Omega_2}
     \cdot \Bigg| \frac{\Omega_1}{\Omega_2} - \frac{m_2}{m_1} \Bigg|^{-1} \,. 
\label{sol:2torus:Psi}
\een 
Now recall that any irrational number, $\Omega_1/\Omega_2 \notin {\Bbb Q}$, 
can be approximated by some rational number $ {m_2/m_1} 
\in {\Bbb Q}$ as close as possible, by taking 
$m_1,\: m_2, \rightarrow \infty$ in an appropriate manner.  
This implies that the denominator of the right side of the above equation can 
become arbitrarily small and therefore that $\Psi$ need not be convergent.

%
Nevertheless, this difficulty has been overcome 
for non-extremal black holes (i.e., the case in which $\kappa \neq 0$,  
see below) by employing a novel approach, and the rigidity result has been 
generalized to higher dimensions by~Refs.~\citen{HIW07,MI08}. 
We quote below the main theorems of Ref.~\citen{HIW07} and briefly describe 
basic ideas of their proof, as well as the attempt of Ref.~\citen{HI09} 
to generalize these theorems to include extremal ($\kappa=0$) black holes.  

\medskip 
\noindent 
{\bf Theorem 3.1:}~[Reference \citen{HIW07}] 
{\em Let $(M,g_{ab})$ be an asymptotically flat, analytic stationary 
black hole solution to the vacuum Einstein equations. 
Assume further that the event horizon, $H$, of the black hole is 
analytic and is topologically $\mr \times \Sigma$, 
with $\Sigma$ compact and connected, and that $\kappa \neq 0$, 
where $\kappa$ is given by Eq.~(\ref{average:surfacegravity}) below.  
Then there exists a Killing field $K^a$, defined in a region that 
covers $H$ and the entire domain of outer communication, such that 
$K^a$ is normal to the horizon and $K^a$ commutes with $t^a$. 
} 

\medskip 
\noindent 
{\bf Theorem 3.2:}~[Reference \citen{HIW07}] 
{\em Under the same assumptions made in Theorem~3.1 above, 
if $t^a$ is not tangent to the generators of $H$, then there exist
mutually commuting Killing fields 
$\varphi^a_{(1)}, \dots, \varphi^a_{(j)}$ ($j\ge 1$) with closed orbits 
with period $2 \pi$ which are defined in a region that covers $H$ 
and the entire domain of outer communication. 
Each of these Killing fields commute with $t^a$, and 
\ben
  t^a = K^a + \Omega_1^{} \varphi_{(1)}^a + \dots 
            +  \Omega_N^{} \varphi_{(N)}^a \, , 
\een
for some constants $\Omega_i$, all of whose ratios are irrational. 
}

\medskip 
Theorem~3.2 implies that if the orbits of $s^a$ fail to be closed, 
then the spacetime must admit at least two linearly independent rotational 
Killing fields.

%
A key new idea employed in Ref.~\citen{HIW07} is to appeal to basic 
results of {\sl von Neumann ergodic theorem}\cite{Walter}---which relies only 
on the compactness of $\Sigma$, and yields that there exists a long-time 
average: 
\ben 
\alpha^*(x) := \lim_{T\rightarrow \infty} 
               \frac{1}{T} \int^T_0 \alpha[\phi_s(x)]{d}s \,. 
\een
Then, using the vacuum Einstein equations, one can show that the 
limit $\alpha^*$ is constant and indeed coincides with the spatial average,  
\ben 
\kappa := \frac{1}{{\rm Area}(\Sigma)} \int_\Sigma \alpha(x) {d}\Sigma\,. 
\label{average:surfacegravity}
\een  
Furthermore, when $\kappa \neq 0$, one can find from the Einstein equations,  
a differential relation $D_a J = \pounds_s \beta_a$ for each of 
the two \eqref{eq:*}, 
with $\beta_a$ being some real analytic vector field on $\Sigma$ 
related to a metric component. 
Then, one can find $\Psi$ as a real analytic solution of 
$D^aD_a\Psi = D_a\beta_a$ on $\Sigma$. 
The ergodic theorem also helps to fix the freedom of adding an solution 
of the homogeneous equations, $D^aD_a \Psi=0$ and $\pounds_s\Psi=0$.

For the extremal black hole case $\kappa=0$, however, 
the above method using the ergodic theory does not seem to apply, 
since, for $\kappa=0$, we do not appear to have 
a differential relation analogous to $D_a J = \pounds_s \beta_a$ 
for one of the two \eqref{eq:*}. 
To proceed, note that in the example of $2$-torus above, 
if for some $q>0$, 
\ben
 \Bigg| \frac{\Omega_1}{\Omega_2} - \frac{m_2}{m_1}  \Bigg|> \frac{1}{m_1^q} 
\,, 
\non 
\een
then one can show that the formal solution, $\Psi$, to 
\eqref{eq:*}, given by Eq.~(\ref{sol:2torus:Psi}), becomes convergent 
and well-defined. 
This condition---called the {\sl Diophantine condition}---does {\sl not} 
hold when $\Omega_1/\Omega_2$ is a Liouville number. However, such a number 
is known to be in a set of {\sl measure zero}. 
Therefore we can virtually always solve Eq.~(\ref{eq:*}).

For the problem of how to solve Eq.~(\ref{eq:*}), we note that 
due to the fact that $s^a$ generates an isometry group on a compact 
Riemannian manifold $\Sigma$, it can locally be decomposed in terms of 
$N\: (\geqslant 1)$ Killing fields $\varphi_i^a$ ($i=1,\dots,N$) as 
\ben
  s^a = \Omega_1 \varphi_1^a+ \cdots + \varphi_N^a \,,  
\label{sadef}
\een
with $\Omega_i \in \mr^N$. Then the following lemma has been 
shown by Ref.~\citen{HI09}. 

\medskip      
\noindent 
{\bf Lemma~3.3:}~[Reference \citen{HI09}] 
{\em Let $J$ be a smooth function on $\Sigma$ with the property 
\ben
   0 = \lim_{T\rightarrow \infty}\frac{1}{T}\int^T_0 J\circ \phi_s d s \,. 
\een
Let $\underline \Omega = (\Omega_1, \dots, \Omega_N) \in \mr^N$ 
satisfy the following ``{\sl Diophantine condition}": 
There exits a number $q$ such that 
\ben
|\bomega \cdot \n| > |\bomega| \cdot |\n|^{-q}
\label{condi:dioph}
\een
holds for all but finitely many $\n \in \mz^N$.
Then the equation \eqref{eq:*} 
with $s^a$ as in Eq.~\eqref{sadef}, has a smooth solution $\Psi$ on $\Sigma$.
Furthermore, if $J$ is real analytic, then the same statements
hold true and $\Psi$ is real analytic. 
} %

\medskip 
With the additional conditions of~\eqref{condi:dioph}, 
the rigidity theorems above have been extended to include 
extremal black holes in Theorems~1 and~2 of Ref.~\citen{HI09}. 
Note that when $N=1$, 
the Diophantine condition is automatically satisfied, and 
when $N>1$---which can happen only in higher dimensions---the condition 
is non-trivial. In this sense, the theorems for the extremal black hole 
case are weaker than the theorems for the non-extremal case. 
However, one also should note that the Diophantine condition holds 
for all $\underline \Omega \in \mr^N$ except for a set of Lebesgue 
measure zero.

\medskip 
A few remarks on the rigidity theorems for both extremal and non-extremal 
cases are in order:   
Theorems~3.1 and~3.2 above, and those corresponding to the extremal case in 
Ref.~\citen{HI09} hold also true for stationary black holes 
coupled to matter fields in a fairly general class of theories 
that include multiple of scalar fields with arbitrary potentials, 
Abelian gauge fields, as well as cosmological constant.  
Thus, the above theorems in particular apply to stationary, 
asymptotically anti-de Sitter black holes as well.

The theorems apply not only to a black hole horizon but also to any horizon 
defined as a ``boundary'' of the causal past of a complete 
orbit of some Killing vector field, such as 
a cosmological horizon if exists.

One can partially remove the analyticity assumption for the black hole 
interior, following the strategy of Refs.~\citen{FRW99,Racz00}. 
For the non-extremal case, the event horizon is isometric to a portion 
of some bifurcate Killing horizon\cite{RW92,RW96}. Then one can use 
the bifurcate horizon as an initial data surface for $K^a$ defined in 
a neighborhood of $H$. Then, applying a characteristic initial value 
formulation to extend $K^a$ into the interior of the black hole.  
This type of characteristic initial value problem is ill-defined 
toward the black hole exterior region and therefore would not appear 
to be applicable to extend $K^a$ into the domain of outer communication.  
Nevertheless, a remarkable progress has recently been made along this 
direction~\cite{IK09a,IK09b}.

The staticity theorem of Sudarsky and Wald\cite{SW92} can be generalized 
straightforwardly to higher dimensions. Then, combined with the rigidity 
theorems, it implies that a stationary, non-extremal black hole 
in $D \geqslant 4$ Einstein-Maxwell system must be either 
static or axisymmetric\cite{Rogatko05}. 
For the former case, we can bypass Steps {\sl (iii)} and {\sl (iv)} of the proof when 
generalizing the uniqueness theorem to higher dimensions as 
in the next section.

\section{Uniqueness of static black holes} 
\label{sec:unique:static}

In this section, we will discuss the uniqueness of static black holes 
in asymptotically flat spacetimes. As seen in chapter 1 and 4 there 
seems to be a large variety of exact solutions with different horizon 
topology, and one cannot restrict the shape of the horizon to be 
spherical in general stationary spacetimes before showing the uniqueness. 
However, we can show the uniqueness of the Schwarzschild-Tangherlini 
spacetimes \cite{Sch} in vacuum static spacetimes, and it is shown that 
the topology is sphere at the same time \cite{Hwang98, GIS1, GIS2}. 

In $4$ dimensions, there are two ways to prove the uniqueness: 
one by Israel\cite{Israel67wq} and the other 
by Bunting \& Masood-ul-Alam\cite{BM}.  
However, both of the methods rely on the special nature of $4$-dimensional 
spacetimes. 
In the former method\cite{Israel67wq}, the Gauss-Bonnet theorem, 
that is, $\int_\Sigma {}^{(2)}\mathscr{R}=8\pi$, is used, where $\Sigma$ is 
a two-dimensional compact surface and ${}^{(2)}\mathscr{R}$ 
is the Ricci scalar. 
In the latter method\cite{BM}, one uses the Weyl-Bach tensor, which vanishes 
if and only if the three dimensional space is conformally flat. 
Therefore, we need a new way to avoid using these four-dimensional 
specialities. 

We first describe some basic formulae for the proof and look at 
the outline of the proof. Then we show the proof of the uniqueness 
theorem for vacuum \cite{Hwang98,GIS1} and electro-vacuum cases \cite{GIS2}.

\subsection{Basic tools}

In this subsection, we describe the basics which will be used for the 
proof of the static black hole uniqueness. The formulae here 
will hold in general static spacetimes. This is because we will not use 
the field equations. 
As commented above, non-rotating black hole spacetimes 
are shown to be static\cite{SW92,Rogatko05}. 
By definition, the staticity means the existence of a timelike Killing 
vector $\xi=\partial /\partial t$ which is hypersurface orthogonal. 
Therefore, we can write down the metric as 
\begin{eqnarray}
ds^2=-V^2(x^i)dt^2+g_{ij}(x^k)dx^i dx^j, \label{static-metric}
\end{eqnarray}
where $\lbrace x^i \rbrace $ are the spatial coordinates. 
Note that the components of the metric do not depend on the time 
coordinate $x^0=t$ and $g_{0i}=0$. 
We denote $t=$constant surface by $S$. 
In this situation, it is natural to consider $(D-1)+1$ decomposition 
of the geometrical quantities.  If black holes exist, they are located 
at $V=0$, that is, the Killing horizon. 

The $D$-dimensional Riemann tensor $R_{MNKL}$ is decomposed into 
\begin{eqnarray}
R_{ijkl}={}^{(D-1)}R_{ijkl} 
\end{eqnarray}
and
\begin{eqnarray}
R_{0i0j}=VD_i D_j V, \label{r0i0j}
\end{eqnarray}
where ${}^{(D-1)}R_{ijkl}$ and $D_i$ are 
the $(D-1)$-dimensional Riemann tensor and 
the covariant derivative with respect to $g_{ij}$, respectively. 
The $D$-dimensional Ricci tensor is decomposed into 
\begin{eqnarray}
R_{00}=VD_i D^i V
\end{eqnarray}
and
\begin{eqnarray}
R_{ij}={}^{(D-1)}R_{ij}-\frac{1}{V}D_i D_jV,
\end{eqnarray}
where ${}^{(D-1)}R_{ij}$ is the $(D-1)$-dimensional 
Ricci tensor. 

The asymptotically flat conditions are 
\begin{eqnarray}
& & V=1-\frac{m}{r^{D-3}}+O(1/r^{D-2}) \label{asym1}\\
& & g_{ij}=\Bigl(1+\frac{2}{D-3}\frac{m}{r^{D-3}} \Bigr)
\delta_{ij} +O(1/r^{D-2}),\label{asym2}
\end{eqnarray}
where  $r:={\sqrt {\delta_{ij} x^i x^j }}$ and $m$ is proportional to 
the ADM mass by 
\begin{eqnarray}
  m={8\pi G\over (D-2) {\rm Vol}(S^{D-2})}M. \label{admmass}
\end{eqnarray}
We can check them by the linear perturbation of the metric. 

Since $V$ often behaves like a monotonic increasing function for 
the outward direction, 
we may employ $V$ as 
a radial coordinate. So we want to introduce the 
unit normal vector as 
\begin{eqnarray}
n_i=\rho D_i V,
\end{eqnarray}
where $\rho$ is the ``lapse" function defined by $\rho:=(D_iV D^iV)^{-1/2}$. 
Then one defines the induced metric orthogonal to $\xi=\partial/\partial t$ 
and $n_i$ as 
\begin{eqnarray}
h_{ij}=g_{ij}-n_in_j.
\end{eqnarray}
The extrinsic curvature of $V=$constant surfaces is defined as 
\begin{eqnarray}
k_{ij}=h_i^k D_k n_j,
\end{eqnarray}
where $h_i^j=g^{jk}h_{ik}$.

To examine the regularity of spacetimes, we often look at the 
Kretschmann invariant
\begin{eqnarray}
R_{IJKL}R^{IJKL}& = & R_{ijkl}R^{ijkl}+4R_{0i0j}R^{0i0j}\nonumber \\
& = & {}^{(D-1)}R_{ijkl}{}^{(D-1)}R^{ijkl}+4V^{-2}D_iD_jVD^iD^j V,
\end{eqnarray}
where we used Eq. (\ref{r0i0j}) in the second line. 
We will rewrite the second term in the second line. 
To do so we go back to the definition of the unit vector 
in the following form
\begin{eqnarray}
D_jV=\rho^{-1}n_j.
\end{eqnarray}
Operating the derivative $D_i$ to the above equation, we have  
\begin{eqnarray}
D_i D_j V=\frac{1}{\rho}k_{ij}-\frac{1}{\rho^2}({\mathscr D}_i \rho n_j+{\mathscr D}_j \rho n_i)
-\frac{1}{\rho^2}n_in_jn^k D_k \rho,
\end{eqnarray}
where ${\mathscr D}_i$ is the covariant derivative with respect to $h_{ij}$. 
Taking the trace of the above equation, we also have the formula
\begin{eqnarray}
k=\rho D_i D^i V+\frac{1}{\rho}n^i D_i \rho.\label{tracek}
\end{eqnarray}
Using the above formulae, the Kretschmann invariant is now rewritten as 
\begin{eqnarray}
R_{IJKL}R^{IJKL}  & = &  
{}^{(D-1)}R_{ijkl}{}^{(D-1)}R^{ijkl} \nonumber \\
& & +\frac{4}{V^2}
\Bigl[ \frac{1}{\rho^2}k_{ij}k^{ij}+\frac{1}{\rho^4} (n^i D_i \rho)^2+
\frac{2}{\rho^4}({\mathscr D}\rho)^2
\Bigr].
\end{eqnarray}
In general, the regularity condition on the horizon ($V=0$) implies 
\begin{eqnarray}
k_{ij}|_{V=0}=n^i D_i \rho |_{V=0}={\mathscr D}_i\rho |_{V=0}=0.
\end{eqnarray}
From Eq. (\ref{tracek}) we see that 
\begin{eqnarray}
D_i D^i V|_{V=0}=0
\end{eqnarray}
holds. Together with the Einstein equation, we will also have the 
constraints on matter fields at the horizon. 

Finally we consider the conformal transformation
\begin{eqnarray}
\tilde g_{ij}= \Omega^2 g_{ij}.
\end{eqnarray}
The Ricci scalar is transformed as 
\begin{eqnarray}
\Omega^2 {}^{(D-1)}\tilde R ={}^{(D-1)}R-2(D-2)D_i D^i 
\ln \Omega-(D-3)(D-2)(D_i \ln \Omega)^2,
\end{eqnarray}
where $(D_i \ln \Omega)^2=D_i \ln \Omega D^i \ln \Omega $. 
The unit normal vector and induced metric are transformed as 
\begin{eqnarray}
\tilde n_i =\Omega n_i
\end{eqnarray}
and
\begin{eqnarray}
\tilde h_{ij}=\tilde g_{ij}-\tilde n_i \tilde n_j=\Omega^2
(g_{ij}-n_in_j)=\Omega^2 h_{ij}.
\end{eqnarray}
For the extrinsic curvature, we see 
\begin{eqnarray}
\tilde k_{ij}=\tilde h_i^k \tilde D_k \tilde n_j 
=\Omega k_{ij}+h_{ij}n^k D_k \Omega.
\end{eqnarray}

In the proof of the uniqueness, we will use the following version of the 
positive mass theorem. 

\vskip 5mm
\noindent {\bf Theorem~4.1:}~[Reference~\citen{PET, PET2}]
{\it Let us consider 
$(D-1)$-dimensional asymptotically flat slices with the non-negative 
Ricci scalar. Then the ADM mass is non-negative. 
Moreover, the slice is flat if and only if the ADM mass vanishes. 
}

\vskip 5mm
Remark: In the above theorem, it is assumed that the slice does not 
have boundaries except for the infinities, and that the metric is $C^1$. 
Although the assumptions for the positive mass theorem are going to be relaxed 
due to various efforts, we would restrict ourself to minimal 
consideration. Because we suppose that the spacetime manifold is 
spin (it is defined so that spinors satisfying a Dirac-type equation 
exist), the positive mass theorem holds for any dimensions $(D \geqslant 4)$.

\subsection{Outline of proof}

There are several different black hole solutions, possessing different types 
of hairs, and detail of the uniqueness proof depends on what type of hairs 
black hole solutions of interest have. 
It is possible to show the uniqueness for a fairly general class of 
static black holes that possess many different hairs. 
However, for such a general case, the proof gets complicated and 
it is hard to see the heart of the proof. For this reason, we will 
first describe the common part of the proof and then go into details 
for a couple of specific cases. 

One often assumes that the outside of black holes is vacuum. 
Once one puts some matter fields or cosmological constant, it becomes 
difficult to show the uniqueness. We also discuss this point shortly. 

Let us look at the outline of the proof. In the proof, we first show that 
the static slice is conformally flat, that is, there is a conformal 
transformation $\tilde g_{ij}=\Omega^2g_{ij}$ so that the resultant space 
is flat ($\tilde g_{ij}=\delta_{ij}$). The conformal flatness is shown 
by the positive mass theorem. More precisely, we can show the 
non-negativity of the Ricci scalar of the conformally transformed 
space and the vanishing of the ADM mass. 
The positive mass theorem implies that the 
space with non-negative Ricci scalar and zero mass must be flat. 
As a result, we realize that the static slice is conformally flat. 
Then we can find a harmonic function, say $u$, (so $\Delta u=0$) 
constructed from the conformal factor $\Omega$ or a part on the flat 
space and the boundary associated with the horizons is spherically symmetric. 
The problem is now equivalent with the electrostatic potential with 
the spherical boundary and we know that each level surface of the potential 
(each $u=$constant surfaces) must be spherically symmetric. 
Finally we can show that the original space $(S, g_{ij})$ is spherically 
symmetric. 
The static and spherically symmetric spacetimes will be a known exact solution 
from the direct calculation or the Birkhoff theorem. 
For the vacuum case, the spacetime is the Schwarzschild-Tangherlini one. 
This is the outline of the proof. 

The main task in the proof is to find an appropriate conformal transformation 
for each theory so that the positive mass theorem can be applied. 
When we know the exact solution, we can guess the conformal transformation 
from the solution itself.

\subsection{Vacuum case}  

Let us first consider the vacuum case. This is the simplest case. 
The vacuum Einstein equation is 
\begin{eqnarray}
R_{MN}=0.
\end{eqnarray}
The black hole solution which we know is the Schwarzschild-Tangherlini 
solution 
\begin{eqnarray}
ds^2=-f(r)dt^2+f(r)^{-1}dr^2+r^2d\Omega_{D-2}^2,
\end{eqnarray}
where 
\begin{eqnarray}
f(r)=1-\frac{2m}{r^{D-3}}, 
\end{eqnarray}
where $m$ is proportional to the ADM mass as Eq.~(\ref{admmass}). 
In the vacuum case, $R_{00}=0$ gives us 
\begin{eqnarray}
D_i D^iV=0.
\end{eqnarray}
Then $V$ is a harmonic function in $t=$constant hypersurfaces $(S,g_{ij})$. 
Since this harmonic function does not have the maximum, we can employ 
$V$ as a radial coordinate on $S$. We will discuss the geometrical feature 
of $V=$constant surfaces (the level surfaces of $V$). 

We employ the following two conformal transformations
\begin{eqnarray}
\tilde g_{ij}^{\pm}=\Omega^2_\pm g_{ij},
\end{eqnarray}
where 
\begin{eqnarray}
\Omega_\pm=\Bigl( \frac{1\pm V}{2} \Bigr)^{\frac{2}{D-3}}
=:\omega_\pm^{\frac{2}{D-3}}.
\end{eqnarray}
Then, from the asymptotic conditions (Eqs. (\ref{asym1}) and (\ref{asym2})), 
we can see the asymptotic behavior of $\tilde g_{ij}^\pm$
at ``infinity"($r = \infty$) 
\begin{eqnarray}
\tilde g_{ij}^+ & = & \Bigl(1-\frac{m}{2r^{D-3}}\Bigr)^{\frac{4}{D-3}}
\Bigl(1+\frac{2}{D-3}\frac{m}{r^{D-3}} \Bigr)\delta_{ij}+O(1/r^{D-2})
\nonumber \\
& = & \delta_{ij}+O(1/r^{D-2})
\end{eqnarray}
and
\begin{eqnarray}
\tilde g_{ij}^- & \simeq & \Bigl(\frac{m}{2r^{D-3}}\Bigr)^{\frac{4}{D-3}}
(dr^2+r^2d\Omega_{D-2}^2) \nonumber \\
& = & (m/2)^{4/(D-3)}r^{-4}(dr^2+r^2d\Omega_{D-2}^2) \nonumber \\
& = & dR^2+R^2d\Omega_{D-2}^2,
\end{eqnarray}
where $R:=(m/2)^{2/(D-3)}r^{-1}$. The reason why we can consider 
two conformal transformation will be seen soon later. 
The mass vanishes in $(\tilde S_+, \tilde g_{ij}^+)$. 
The infinity $r=\infty$ in $(S, g_{ij})$ corresponds 
to be a point 
in $(\tilde S_-, \tilde g_{ij}^-)$. 
The Ricci scalar of $\tilde g_{ij}^\pm$ becomes 
\begin{eqnarray}
{}^{(D-1)} \tilde R_\pm=0.
\end{eqnarray}

Let us try to construct the manifold 
$(\tilde S, \tilde g_{ij}):=
(\tilde S_+, \tilde g_{ij}^+) \cup 
(\tilde S_-, \tilde g_{ij}^-) \cup 
\lbrace p \rbrace$ gluing along the $V=0$ surfaces and adding the 
point $p$. $(\tilde S_-, \tilde g_{ij}^-) \cup \lbrace p\rbrace$ buries the 
``hole" of $(\tilde S_+, \tilde g_{ij}^+)$. 
On the horizon, it is easy to see that the induced metric in 
$(\tilde S_\pm, \tilde g_{ij}^\pm)$ is continuous, 
that is, 
\begin{eqnarray}
\tilde h_{ij}^+|_{V=0}=\tilde h_{ij}^-|_{V=0}.
\end{eqnarray}
The extrinsic curvature of the $V=0$ surface in $(\tilde S_\pm, \tilde g_{ij}^\pm)$
has the same absolute value with opposite signature
\begin{eqnarray}
\tilde k_{ij}^\pm |_{V=0}=\pm \frac{2^{\frac{D-1}{D-3}}}{D-3} 
\frac{1}{\rho_0}\tilde h_{ij}^\pm |_{V=0}.
\label{totally}
\end{eqnarray}
The surface satisfying the relation (\ref{totally}) 
is said to be totally umbilical. 
The difference of the signature comes from that of the 
normal direction to the $V=0$ surfaces. Since the extrinsic curvature is 
expressed by the Lie derivative as 
\begin{eqnarray}
\tilde k_{ij}=\frac{1}{2} \pounds_{\tilde n}\tilde h_{ij},
\end{eqnarray}
the metric of $\tilde g_{ij}$ is $C^1$. And 
$\tilde S$  does not have the boundary except for the infinities. 

Then we can apply the positive mass 
theorem to $(\tilde S, \tilde g_{ij})$ and show that 
$(\tilde S, \tilde g_{ij})$ is flat, 
\begin{eqnarray}
\tilde g_{ij}^\pm=\delta_{ij}. 
\end{eqnarray}
We also note that the totally umbilical surfaces in the flat space is 
spherically symmetric \cite{KN5-1}. 

The equation for $V$, $D^2V=0$, is rewritten in terms of $g_{ij}$ as 
follows
\begin{eqnarray}
\tilde D^i [ \Omega^{-(D-3)}_\pm \tilde D_i V]=0.
\end{eqnarray}
Then, defining a function $u_\pm$ by 
\begin{eqnarray}
u_\pm=\frac{1}{\Omega_\pm}=\frac{2}{1\pm V},
\end{eqnarray}
we see that 
\begin{eqnarray}
\tilde D_i \tilde D^i u_\pm=\Delta_0u_\pm=0
\end{eqnarray}
holds, where $\Delta_0$ is the Laplacian of the flat space. 

Now the problem here can be reduced to the electrostatic potential with 
the spherical boundary in the flat space. Then it is easy to see that 
the $u_\pm$=constant surfaces are spherically symmetric in 
$(\tilde S_+, 
\tilde g_{ij}^+)$. Since $u_\pm$ and 
$\Omega_\pm$ are the functions of $V$ only, $V=$constant surfaces are also 
spherically symmetric in $(S, g_{ij})$. Through the usual computation, 
it is shown that a static, spherically symmetric vacuum black hole 
solution is isometric to the Schwarzschild-Tangherlini 
metric. This is the end of the proof. 

Before closing this subsection, we will express the spacetimes in the 
isotropic coordinates to guess the conformal transformation for the 
proof of the uniqueness. This may give us a hint for other 
black holes. In the isotropic coordinates, the metric of 
the Schwarzschild-Tangherlini spacetime is 
\begin{eqnarray}
ds^2=-\frac{\Bigl(1-\frac{m}{2\rho^{D-3}} \Bigr)^2}{\Bigl(1+\frac{m}{2\rho^{D-3}} \Bigr)^2}dt^2
+\Bigl(1+\frac{m}{2\rho^{D-3}} \Bigr)^{\frac{4}{D-3}}(d\rho^2+\rho^2 d\Omega_{D-2}^2), 
\end{eqnarray}
where the relation between $r$ and $\rho$ is given by 
\begin{eqnarray}
r=\rho \Bigl(1+\frac{m}{2\rho^{D-3}} \Bigr)^{\frac{2}{D-3}}. 
\end{eqnarray}

Compared to the general form of the static spacetime 
(Eq.~(\ref{static-metric})), we can identify $V$ as 
\begin{eqnarray}
V=\frac{1-\frac{m}{2\rho^{D-3}}}{1+\frac{m}{2\rho^{D-3}}}.
\end{eqnarray}
Then one of the conformal transformations is expected to be 
\begin{eqnarray}
\Omega_+=\Bigl(1+\frac{m}{2\rho^{D-3}} \Bigr)^{-\frac{2}{D-3}}
=\Bigl( \frac{1+V}{2} \Bigr)^{\frac{2}{D-3}}.
\end{eqnarray}
Introducing the new coordinate defined by 
\begin{eqnarray}
R:=\rho^{-1}(m/2)^{\frac{2}{D-3}},
\end{eqnarray}
the spatial part of the metric is rewritten as 
\begin{eqnarray}
d\ell^2=\Bigl(1+\frac{m}{2R^{D-3}} \Bigr)^{\frac{4}{D-3}}(dR^2+R^2d\Omega_{D-2}^2).
\end{eqnarray}
In this coordinate, $\rho = \infty$ corresponds to $R=0$. 
Then we can expect the other conformal transformation to be 
\begin{eqnarray}
\Omega_-=\Bigl(1+\frac{m}{2R^{D-3}} \Bigr)^{-\frac{2}{D-3}}=
\Biggl[\frac{\rho^{D-3}}{m/2}\Bigl(1+\frac{m}{2\rho^{D-3}} \Bigr) \Biggr]^{-\frac{2}{D-3}}
=\Bigl( \frac{1-V}{2} \Bigr)^{\frac{2}{D-3}}.
\end{eqnarray}
They are exactly the same as those employed in the proof of the 
uniqueness theorem. 

\subsection{Electro-vacuum case}

Next we will consider charged black hole solutions. The Lagrangian is 
given by 
\begin{eqnarray}
16\pi G L=R-F_{MN}F^{MN},
\end{eqnarray}
where $F_{MN}$ is the field strength of the Maxwell fields. The 
Einstein equation and Maxwell equation are 
\begin{eqnarray}
R_{MN}=2F_{M}^{~K}F_{NK}-\frac{1}{D-2}g_{MN}F^2 \label{em}
\end{eqnarray}
and
\begin{eqnarray}
\nabla_M F^{MN}=0
\end{eqnarray}
with $\nabla_{[M}F_{NI]}=0$. 
In general static spacetimes, the electric part of the Maxwell field 
becomes
\begin{eqnarray}
F_{0i}=-\partial_i \psi (x^j). 
\end{eqnarray}
Here we note that one may be interested in the higher form fields because 
they are essential ingredient in string theory. 
However, the possible global charge is 
only for the Maxwell field. This is because to compute the global charge, 
we need to perform an integration of the field strength over a sphere with 
infinite radius. To specify the sphere in spacetimes, we need 
two directions which are orthogonal to the sphere. 
Therefore, we cannot define the global charge of the higher form field 
strength. We will comment on the hair of higher form fields 
in the final subsection again. 

The solution which we know is the Reissner-Nordstr\"{o}m solution 
\begin{eqnarray}
ds^2=-f(r)dt^2+f(r)^{-1}dr^2+r^2d\Omega_{D-2}^2,
\end{eqnarray}
where 
\begin{eqnarray}
f(r)=1-\frac{2m}{r^{D-3}}+\frac{Q^2}{r^{2(D-3)}}. 
\end{eqnarray}
The electric potential $\psi$ is given by 
\begin{eqnarray}
\psi=\frac{Q/C}{r^{D-3}}, 
\end{eqnarray}
where
\begin{eqnarray}
C:=\Bigl(\frac{2(D-3)}{D-2} \Bigr)^{1/2}. 
\end{eqnarray}
We shall show the uniqueness of the Reissner-Nordstr\"{o}m solution in the 
Einstein-Maxwell theory. In this subsection we assume $m>|Q|$. We will 
briefly comment on the extreme case of $m=|Q|$ in the final subsection. 

From Eq. (\ref{em}), we have 
\begin{eqnarray}
R_{00}=VD_i D^i V=C^2(D_i \psi)^2,
\end{eqnarray}
and
\begin{eqnarray}
R_{ij}={}^{(D-1)}R_{ij}-\frac{1}{V}D_i D_j V=
-\frac{2}{V^2}D_i \psi D_j \psi+\frac{2}{D-2}g_{ij}\frac{1}{V^2}
(D_k \psi)^2.
\end{eqnarray}
The Maxwell equation becomes
\begin{eqnarray}
D_i D^i \psi=\frac{D^iV D_i \psi}{V}.
\end{eqnarray}
The asymptotic condition for $\psi$ is given by 
\begin{eqnarray}
\psi=\frac{Q/C}{r^{D-3}}+O(1/r^{D-2}).
\end{eqnarray}

Let us consider the following conformal transformation
\begin{eqnarray}
\tilde g_{ij}=\Omega_\pm^2 g_{ij},
\end{eqnarray}
\begin{eqnarray}
\Omega_\pm=\Bigl[\Bigl( \frac{1\pm V}{2}\Bigr)^2-\frac{C^2}{4}\psi^2 \Bigr]^{\frac{1}{D-3}}
=:\omega_\pm^{\frac{1}{D-3}}=(\lambda_\pm \mu_\pm)^{\frac{1}{D-3}},
\end{eqnarray}
where 
\begin{eqnarray}
\lambda_\pm=\frac{1\pm V}{2}-\frac{C}{2}\psi
\end{eqnarray}
and
\begin{eqnarray}
\mu_\pm=\frac{1\pm V}{2}+\frac{C}{2}\psi.
\end{eqnarray}
From the later discussion about the exact solution, we will see why 
they are chosen. The asymptotic behavior of $\tilde g_{ij}^\pm$ is 
\begin{eqnarray}
\tilde g_{ij}^+=\delta_{ij}+O(1/r^{D-2})
\end{eqnarray}
and
\begin{eqnarray}
\tilde g_{ij}^- \simeq dR^2+R^2d\Omega_{D-2}^2,
\end{eqnarray}
where 
\begin{eqnarray}
r^{-1}=\Bigl(\frac{m^2-Q^2}{4} \Bigr)^{\frac{1}{D-3}}R.
\end{eqnarray}
As in the vacuum case, the ADM mass in $(\tilde S_+, \tilde g_{ij}^+)$ 
vanishes and the infinity $(r=\infty)$ becomes to be a point $p$ in 
$(\tilde \Sigma_-, \tilde g_{ij}^-)$. 

From the Einstein equation, the $(D-1)$-dimensional Ricci scalar is 
computed as  
\begin{eqnarray}
{}^{(D-1)}R=2\frac{(D_i \psi)^2}{V^2}. 
\end{eqnarray}
Then the Ricci scalar of $\tilde g_{ij}^\pm $ becomes 
\begin{eqnarray}
\Omega_\pm^2 {}^{(D-1)}\tilde R_\pm
& = & 2 \frac{(D_i\psi)^2}{V^2}
-\frac{2(D-2)}{(D-3)}\Bigl( \frac{D_i D^i\lambda_\pm}{\lambda_\pm}
+\frac{D_i D^i\mu_\pm}{\mu_\pm}\Bigr) \nonumber \\
& & +\frac{D-2}{D-3}\Bigl(\frac{D_i\lambda_\pm}{\lambda_\pm}
-\frac{D_i\mu_\pm}{\mu_\pm} \Bigr)^2.
\end{eqnarray}
Here we note that, for $\lambda_\pm$ and $\mu_\pm$, the following equations 
hold 
\begin{eqnarray}
& & D_i D^i \lambda_\pm=\mp \frac{C}{V}D^i\psi  D_i \lambda_\pm \label{lambda}\\
& & D_i D^i \mu_\pm=\pm \frac{C}{V}D^i\psi D_i \mu_\pm. \label{mu}
\end{eqnarray}
Using them, the Ricci scalar of $\tilde g_{ij}^\pm$ 
can be expressed as 
\begin{eqnarray}
\Omega^2_\pm {}^{(D-1)}\tilde R_\pm & = & \frac{2}{C^2}
\Bigl(\frac{D_i \lambda_\pm}{\lambda_\pm}-\frac{D_i \mu_\pm}{\mu_\pm}
\pm C \frac{D_i \psi}{V} \Bigr)^2 \nonumber \\
& = & \frac{C^2}{8\omega_\pm^2V^2}
\Biggl[ 
2\psi VD_iV+(1-V^2-C^2\psi^2)D_i\psi\Biggr]^2. 
\end{eqnarray}
Then we can see 
\begin{eqnarray}
{}^{(D-1)}\tilde R_\pm \geq 0.
\end{eqnarray}

The regularity on the horizon implies 
\begin{eqnarray}
k_{ij}|_{V=0}={\mathscr D}_i \rho|_{V=0}=D_i \psi|_{V=0}=0 . 
\end{eqnarray}
Then the induced metric and the extrinsic curvature of $V=0$ surface in 
$(\tilde S, \tilde g_{ij})$ become
\begin{eqnarray}
& & \tilde h_{ij}^+|_{V=0}= \tilde h_{ij}^-|_{V=0} \\
& & \tilde k_{ij}|_{V=0}=
\pm \frac{2^{\frac{D-1}{D-3}}}{(D-3)(1-C^2\psi_0^2)^{\frac{D-2}{D-3}}}
\frac{1}{\rho_0}\tilde h_{ij}^\pm|_{V=0}.
\end{eqnarray}

Let us construct the manifold 
$(\tilde S, \tilde g_{ij}):=(\tilde S_+,\tilde g_{ij}^+)
\cup (\tilde S^-, \tilde g_{ij}^-)\cup \lbrace p \rbrace$. 
As in the vacuum case, the metric and extrinsic curvature are continuous. 
Thus we can apply the positive mass theorem to 
$(\tilde S, \tilde g_{ij})$. Since $(\tilde S, \tilde g_{ij})$ 
is massless and the Ricci scalar is non-negative, 
the positive mass theorem implies that $(\tilde S, \tilde g_{ij})$ 
must be flat, that is, 
\begin{eqnarray}
\tilde g_{ij}=\delta_{ij} . 
\end{eqnarray}
The expression of ${}^{(D-1)}\tilde R_\pm$ gives us 
\begin{eqnarray}
2V\psi D_i V+(1-V^2-C^2\psi^2)D_i \psi=0. 
\end{eqnarray}
This means that the level surface of $V$ is equivalent with 
that of $\psi$. 

Now we can realize that Eqs. (\ref{lambda}) and (\ref{mu}) are 
rewritten in terms of $\tilde g_{ij}$ as 
\begin{eqnarray}
\lambda_\pm^2 \tilde D^2 \lambda_\pm^{-1}
=-\tilde D^i \lambda_\pm 
\Bigl(\tilde D_i \ln \frac{\mu_\pm}{\lambda_\pm}\mp 
\frac{C}{V}\tilde D_i \psi  \Bigr)=0
\end{eqnarray}
and
\begin{eqnarray}
\mu_\pm^2 \tilde D^2 \mu_\pm^{-1}
=\tilde D^i \mu_\pm 
\Bigl(\tilde D_i \ln \frac{\mu_\pm}{\lambda_\pm}\mp 
\frac{C}{V}\tilde D_i \psi  \Bigr)=0.
\end{eqnarray}
Therefore, $\lambda_\pm^{-1}$ and $\mu_\pm^{-1}$ are harmonic functions 
in $(\tilde S_\pm, \tilde g_{ij}^\pm)$. Each level surface is the same. 
Then we take 
$\lambda^{-1}_+$ to show the spherical symmetry. As in the vacuum case, 
$V=0$ surfaces in $(\tilde S, \tilde g_{ij})$ is totally 
umbilical and this means that the $V=0$ surface is spherically symmetric 
in the flat space. 
Thus, the problem is reduced to be the electro-static potential 
one with spherically symmetric boundary. So the level surfaces of 
$\lambda_+^{-1}$ are all spherically symmetric in 
$(\tilde S_+, \tilde g_{ij}^+)$. 
Since $\Omega_+=\Omega_+(V,\psi)$ and the level surfaces of $V$ are 
the same as those of $\psi$, $g_{ij}=\Omega_+^{-2}\delta_{ij}$ is also 
spherically symmetric. 
Solving the Einstein-Maxwell equations directly, we can see that the 
solution is isometric to the Reissner-Nordstr\"{o}m metric. 
This is the end of the proof. 

At first glance the choice of the conformal transformation is non-trivial 
although it is quite important. The choice can be understood from the 
Reissner-Nordstr\"{o}m solution in the isotropic coordinates
\begin{eqnarray}
ds^2& = & -\Biggl( 
\frac{1-\frac{m^2-Q^2}{4}\frac{1}{\rho^{2(D-3)}}}{1+\frac{m}{\rho^{D-3}}+\frac{m^2-Q^2}{4}\frac{1}{\rho^{2(D-3)}}} 
\Biggr)^2dt^2 \nonumber \\
& & +\Bigl(1+\frac{m}{\rho^{D-3}}+\frac{m^2-Q^2}{4}\frac{1}{\rho^{2(D-3)}}\Bigr)^{\frac{2}{D-3}}
(d\rho^2+\rho^2 d\Omega_{D-2}^2),
\end{eqnarray}
where $\rho$ is introduced through the relation of 
\begin{eqnarray}
r=\rho \Bigl(1+\frac{m}{\rho^{D-3}}+\frac{m^2-Q^2}{4}\frac{1}{\rho^{2(D-3)}} \Bigr)^{\frac{1}{D-3}}.
\end{eqnarray}
The electric potential is 
\begin{eqnarray}
\psi=\frac{Q}{\rho^{D-3}}\frac{1}{1+\frac{m}{\rho^{D-3}}+\frac{m^2-Q^2}{4}\frac{1}{\rho^{2(D-3)}}}. 
\end{eqnarray}
Compared to the metric form of general static spacetimes, we can identify 
$V$ as 
\begin{eqnarray}
V=\frac{1-\frac{m^2-Q^2}{4}\frac{1}{\rho^{2(D-3)}}}{1+\frac{m}{\rho^{D-3}}+\frac{m^2-Q^2}{4}\frac{1}{\rho^{2(D-3)}}}. 
\end{eqnarray}
Since 
\begin{eqnarray}
\Bigl( \frac{1+V}{2}\Bigr)^2-\frac{C^2}{4}\psi^2
=\frac{1}{1+\frac{m}{\rho^{D-3}}+\frac{m^2-Q^2}{4}\frac{1}{\rho^{2(D-3)}}},
\end{eqnarray}
we may choose the conformal transformation 
\begin{eqnarray}
\Omega_+
& = & 
\Bigl(1+\frac{m}{\rho^{D-3}}+\frac{m^2-Q^2}{4}\frac{1}{\rho^{2(D-3)}} 
\Bigr)^{-\frac{1}{D-3}} \nonumber \\
& = & \Biggl[\Bigl( \frac{1+V}{2}\Bigr)^2-\frac{C^2}{4}\psi^2 \Biggr]^{\frac{1}{D-3}}.
\end{eqnarray}

Introducing the new coordinate 
\begin{eqnarray}
R:=\Bigl( \frac{m^2-Q^2}{4}\Bigr)\rho^{-1},
\end{eqnarray}
the spatial part of the metric is written as 
\begin{eqnarray}
d\ell^2
=\frac{\rho^4}{\Bigl( \frac{m^2-Q^2}{4}\Bigr)^{\frac{2}{D-3}}}
\Bigl(1+\frac{m}{\rho^{D-3}}+\frac{m^2-Q^2}{4}\frac{1}{\rho^{2(D-3)}}
\Bigr)^{\frac{2}{D-3}}(dR^2+R^2d\Omega_{D-2}^2). 
\end{eqnarray}
Note that 
\begin{eqnarray}
\Bigl( \frac{1-V}{2}\Bigr)^2-\frac{C^2}{4}\psi^2
=\frac{\frac{m^2-Q^2}{4}}{\rho^{2(D-3)}}
\frac{1}{1+\frac{m}{\rho^{D-3}}+\frac{m^2-Q^2}{4}\frac{1}{\rho^{2(D-3)}}} 
\end{eqnarray}
holds. They also give us a hint for another choice 
of the conformal transformation, that is, 
\begin{eqnarray}
\Omega_-
& = & 
\Biggl[\frac{\rho^{2(D-3)}}{\frac{m^2-Q^2}{4}}
\Bigl(  1+\frac{m}{\rho^{D-3}}+\frac{m^2-Q^2}{4}\frac{1}{\rho^{2(D-3)}} \Bigr)  \Biggr]^{-\frac{1}{D-3}}
\nonumber \\
& = & \Biggl[\Bigl( \frac{1-V}{2}\Bigr)^2-\frac{C^2}{4}\psi^2 \Biggr]^{\frac{1}{D-3}}.
\end{eqnarray}

\subsection{Other cases}

From the view point of simpleness, we have considered the vacuum and 
electrovacuum cases. There are also uniqueness theorems for other 
static black holes. We will briefly comment on each case. 

\smallskip \noindent  
{\sl (i) Dilatonic black holes}: In higher dimensions, we have the exact 
solution of the dilatonic Einstein-Maxwell theory \cite{Gibbons:1987ps}. 
In this case, we can also prove the uniqueness in the same manner 
here \cite{GIS2}. The same technique also applies to the case 
coupled to sigma-model fields\cite{Rogatko02}. 

\smallskip \noindent  
{\sl (ii) No-dipole hair theorem for form fields}: One may be interested 
in the hair of the $p$-form field strength of 
black holes. When one can show the no-hair, it is, of course, 
expected that exact solutions with such hair cannot exist. 
Therefore, we cannot have the hint for the conformal transformation 
from solutions because there are no known hairy solutions. 
For $p\geqslant 3$, possible 
hairs are dipole or higher multipole components of the field 
strengths. However, we can show the no-hair of the electric $p$-form field 
strength when $p\geqslant (D+1)/2$ \cite{EOS}. 
In the proof of this theorem, we use the same conformal transformation 
employed in the proof of the Schwarzschild-Tangherlini spacetime. 

\smallskip \noindent  
{\sl (iii) Extreme black holes in Einstein-Maxwell theory}: 
In the previous subsection, the non-extreme condition $m>|Q|$ 
is imposed. In the extreme limit, there are exact solutions  
for multi-black holes. The uniqueness of such solutions has been proven 
\cite{Chrusciel:2005pa,Rogatko03}. See Ref.~\citen{Rogatko06} for some
related work.  

\smallskip \noindent  
{\sl (iv) Asymptotically de Sitter/anti-de Sitter}: 
There are exact solutions of the Schwarzschild-(anti-)de Sitter 
solutions. 
See Refs.~\citen{Boucher:1983cv,ACD02} for attempts to generalize 
the static uniqueness to the asymptotically anti-de Sitter case.
See also Ref.~\citen{Kodama.H2004} for perturbative approach.

\vskip 5mm
There are also asymptotically flat supersymmetric black hole solutions 
in four and five dimensions. Since they are 
extreme type, they will be multi-black hole solutions. In 
four dimensions, there is a sort of uniqueness theorem \cite{Chrusciel:2005pa}. 
In five dimensions, the uniqueness of the near-horizon geometry is 
discussed \cite{Reall03}.

\section{Uniqueness of stationary, rotating black holes} 
\label{sec:unique:stationary}

\subsection{Basic strategy} 
\label{subsec:unique:stationary:strategy}
Now we turn to stationary, rotating vacuum black holes. 
A critical step toward the uniqueness proof is the reduction of the problem 
to a boundary value problem of a certain type of non-linear sigma model 
[recall Step {\sl (iii)} of section~\ref{sec:stationary}]. 
In $4$-dimensions, the rigidity theorem applying a rotational black hole 
ensures the existence of the $2$-dimensional isometry group, 
$\mr \times U(1)$, which is enough to reduce the system to a sigma model 
on the $2$-dimensional factor space $M/\{\mr \times U(1)\}$, 
thereby being decoupled from gravity. 
In higher dimensions, the rigidity theorem again holds as seen previously, 
but guarantees the existence of only one rotational symmetry. 
Therefore, in order to take the same steps as in $4$-dimensions, 
we need to assume multiple axial symmetries, $U(1)^{D-3}$, besides 
the stationarity $\mr$.

%
Another important step is to identify boundary conditions on the sigma model 
fields that are necessary and sufficient to fully determine the solution 
[recall Step {\sl (iv)} of section~\ref{sec:stationary}].  
In $4$-dimensions, the desired boundary conditions are given by 
specifying asymptotic conserved charges, such as the total mass and angular 
momentum. 
However, in higher dimensions, these conditions are not enough and we need 
to specify more data than merely having the same asymptotic conserved charges. 
We at least need to specify the topology of a black hole considered. 
As the phase diagram of $5$-dimensional stationary rotating vacuum black holes 
indicates, if we restrict attention to spherical black holes, 
then the $4$-dimensional uniqueness result may be generalized to 
$5$-dimensions. 
Such a restricted case of uniqueness theorems has been shown 
for the first time by Morisawa and Ida\cite{Morisawa:04}.  
We shall discuss how to prove the uniqueness theorem for topologically 
restricted, spherical horizon case in the next subsection. 
But before going into that, let us illustrate basic strategy for 
the uniqueness proof in more general context.

According to Steps {\sl (iii)} and {\sl (iv)},  we proceed as follows. 
(1) We first reduce the $D$-dimensional Einstein equations with 
$(D-2)$ commuting Killing vector fields (one of which is 
for the stationarity) to a non-linear sigma model, 
that is, a set of equations for scalar fields 
on the $2$-dimensional factor space $B$, with the target space isometry $G$. 
With the aid of $G$, the action of the sigma model can be described 
in terms of a symmetric, unimodular matrix, $\Phi$, on the coset space $G/H$ 
where $H$ is an isotropy subgroup of $G$. Thus, the solutions of the original 
system can be expressed by the matrix $\Phi$. Furthermore, the matrix 
$\Phi$ formally defines a conserved current, $\Pi_i$, for the solution. 
(2) Next, we introduce the {\sl deviation matrix}, $\Psi$, which is  
essentially the difference between two coset matrices, 
say $\Phi$ and $\Phi'$, so that when the two solutions coincide with 
each other, the deviation matrix vanishes, and vice versa;  
it is usually defined as $\Psi :=\Phi'\Phi^{-1}-I$.  
What we wish to show is that $\Psi$ vanishes over the entire $B$ when the 
two solutions satisfy the same boundary conditions that specify relevant 
physical parameters characterising the black hole solution of interest. 
For this purpose, we construct a global identity, called the 
{\sl Mazur identity}, (the integral version of) which equates an integration 
along the boundary $\partial B$ of a derivative of the trace of $\Psi$ 
to an integration over the whole base space $B$ of the trace of 
`square' of the deviation, $M_i$, of the two conserved currents, 
$\Pi_i$ and $\Pi_i'$. The latter is therefore non-negative. 
(3) Then, we perform boundary value analysis of the matrix $\Psi$.   
We identify boundary conditions for $\Phi$ (and $\Phi'$) that define 
physical parameters characterising the corresponding black hole solutions 
and that guarantee the regularity of the solutions. 
Then we examine the behavior of $\Psi$ near $\partial B$. 
For higher dimensional case, this is the point where we need, 
other than the asymptotic conserved charges, some additional parameters 
to fully specify the solutions. 
Also this is the place where we have to take into consideration
the nature of asymptotic structure of the spacetime. 
When the integral along the boundary $\partial B$, say the left-side 
of the Mazur identity, vanishes under our boundary conditions, it then 
follows from the right-side of the identity, i.e., the non-negative 
integration over $B$, that $M_i$ has to vanish, hence 
the two currents, $\Pi_i$ and $\Pi'_i$, must coincide with each other 
over $B$, 
implying that the deviation matrix $\Psi$ must be constant over $B$. 
Then, if $\Psi$ is shown to be zero on some part of the boundary 
$\partial B$, it follows that $\Psi$ must be identically zero 
over the entire $B$, thus proving the two solutions, 
$\Phi$ and $\Phi'$, must be identical. 

\subsection{Rotating black holes in 5-dimensions}  
\label{subsec:unique:stationary:5D}
From now on we focus on the asymptotically weakly simple $5$-dimensional 
spacetimes admitting Abelian isometry group 
$\boldsymbol{R}\oplus U(1)\oplus U(1)$. 
The isometry group is generated by three commuting Killing vector fields 
$U$, $V$ and $W$, where $U$ denotes the generator of the
time translation, and each $V$ and $W$ generates spatial rotation
with respect to one of the pair of half 2-planes (namely, the symmetric axes), 
whose orbits are circles. 
This requirement of isometry is consistent with the asymptotic flatness, 
for $U(1)\oplus U(1)$ is contained in the rotational group 
$SO(4)\simeq SU(2)\times SU(2)$ as a subgroup. 

The Frobenius conditions
\begin{eqnarray}
(\partial_{[a} U_b) U_cV_dW_{e]}&=&0,\nonumber\\
(\partial_{[a} V_b) U_cV_dW_{e]}&=&0,\nonumber\\
(\partial_{[a} W_b) U_cV_dW_{e]}&=&0,\nonumber
\end{eqnarray}
 for the integrability of the
2-dimensional distribution orthogonal to the group orbits of the isometry
is contained in the vacuum Einstein equation.
This can be seen from the identities
\begin{eqnarray}
\partial_f [(-g)^{1/2}\epsilon_{abcde}U^aV^bW^c\nabla^dU^e]
&=&2(-g)^{1/2}\epsilon_{abcdf}U^aV^bW^cR^d_eU^e,
\nonumber\\
\partial_f [(-g)^{1/2}\epsilon_{abcde}U^aV^bW^c\nabla^dV^e]
&=&2(-g)^{1/2}\epsilon_{abcdf}U^aV^bW^cR^d_eV^e,
\nonumber\\
\partial_f [(-g)^{1/2}\epsilon_{abcde}U^aV^bW^c\nabla^dW^e]
&=&2(-g)^{1/2}\epsilon_{abcdf}U^aV^bW^cR^d_eW^e.
\nonumber
\end{eqnarray}
From the Ricci flatness, the right hand side of each identity becomes zero. This implies that
each quantity between the square brackets in the left hand side is constant.
However, all these quantities are zero at the fixed points of the symmetric 
axes, so they must be zero everywhere. This clearly shows the fulfillment of 
the Frobenius conditions. 

The integrability of the 2-distribution orthogonal to the orbits of 
the isometric actions implies that
there is a coordinate system
$(x^1,\cdots,x^5)=(x,y,t,\phi,\psi)$, 
where the Lorentzian metric is written in the block diagonal form
\begin{eqnarray}
g=\sum_{i,j=1}^2h_{ij}(x,y)dx^idx^j+\sum_{I,J=3}^5\Phi_{IJ}(x,y)dx^Idx^J,\nonumber
\end{eqnarray}
and the Killing vector fields are expressed as
\begin{eqnarray}
  U&=&\partial_t,\quad
  V=\partial_\phi,\quad
  W=\partial_\psi.
\end{eqnarray}

We mean the asymptotic flatness by imposing that the metric has an asymptotic form
\begin{eqnarray}
g&=&-\left[1-{\frac{8GM}{3\pi r^2}}+O(r^{-3})\right]dt^2
+\left[1+{4GM\over 3\pi r^2}+O(r^{-3})\right](dx^2+dy^2+dz^2+dw^2)
\nonumber
\\
&&-\left[{8GJ_\phi\over \pi r^4}+O(r^{-5})\right]dt(ydx-xdy)
-\left[{8GJ_\psi\over \pi r^4}+O(r^{-5})\right]dt(wdz-zdw),
\label{AF}
\end{eqnarray}
where $M$, $J_\phi$ and $J_{\psi}$ denote the ADM mass and angular momenta, respectively.
We further assume that 
\begin{eqnarray}
  {32GM^3\over 27\pi}>(|J_\phi|+|J_\psi|)^2\nonumber
\end{eqnarray}
holds.

At this stage, the gravitational fields are described by the 6 scalar fields $\{\Phi_{IJ}(x,y)\}$
and the Riemannian metric $h(x,y)$ on the 2-dimensional base space $B^2$ parametrized
by $(x,y)$. More convenient parametrization of $\{\Phi_{IJ}\}$ 
and $h$
is given by
\begin{eqnarray}
\sum_{I,J=3}^5\Phi_{IJ}dx^Idx^J&=&
\sum_{I,J=4}^5f_{IJ}(dx^I+w^Idt)(dx^J+w^Jdt)-f^{-1}\rho^2 dt^2,\\
h&=&f^{-1}\gamma,
\label{gram:matrix}
\end{eqnarray}
where $\rho^2=-{\rm det}(\Phi_{IJ})\ge 0$ and $f={\rm det}(f_{IJ})$.
Note that the function $\rho^2$ becomes zero only on the symmetric axes, 
which are actually $2$-planes though we informally call them axes, 
and on the event horizon.
Furthermore, we introduce the twist potentials $\omega_I$ $(I=4,5)$ by
\begin{eqnarray}
  \partial_k\omega_I=\rho^{-1}f  f_{IJ}\sqrt{\gamma}
\epsilon_{ik} \gamma^{im}\partial_m w^J,\label{def_twist}
\end{eqnarray}
where $\epsilon_{ik}$ is the antisymmetric symbol on the 2-space $B^2$
specified by $\epsilon_{12}=1$, and $\gamma={\rm det}(\gamma_{ij})$.
The integrability condition of Eq.~(\ref{def_twist})
is provided by the Ricci flat condition via
\begin{eqnarray}
  \partial_{[i}\partial_{j]}\omega_I
=\rho f^{-1}\sqrt{\gamma}R_I^3\epsilon_{ij}.\nonumber
\end{eqnarray}
Hence the twist potential $\omega_I$ is globally defined 
if the 2-dimensional base space $B^2$ is simply connected as assumed later. 

The Ricci flat condition also ensures that the function $\rho$ is a harmonic 
function on the base space. This can be seen from the identity 
\begin{eqnarray}
  \Delta^\gamma\rho=\rho^{-1}R_{33}-\rho f^{-1}f^{IJ}R_{IJ},
\end{eqnarray}
where $\Delta^\gamma$ denotes the Laplace operator on $B^2$, and $f^{IJ}$ is 
the component of the inverse matrix of $f_{IJ}$.
We choose the harmonic function  $\rho\ge 0$ and its conjugate harmonic 
function $z$ as coordinates on $B^2$, so that the pair $(\rho,z)$ gives 
an isothermal coordinate system. 
[See Ref.~\citen{Chr09} for a rigorous proof of the well-definedness 
of $\rho$,  which is based on the results of Refs.~\citen{CC08,CGS09}.
See also earlier work~\citen{Carter71}.]
Hence, the spacetime metric can be written as 
\begin{eqnarray}
  g=f^{-1}e^{2\sigma}(d\rho^2+dz^2)-f^{-1}\rho^2 dt^2
+\sum_{I,J=4}^5f_{IJ}(dx^I+w^Idt)(dx^J+w^Jdt),\nonumber
\end{eqnarray}
where all the metric functions depend only on $\rho$ and $z$. 
This form of the metric is often called the {\sl Weyl-Papapetrou type form}. 

Now, the Einstein equation is reduced to
the elliptic equations on the $(\rho,z)$-half plane 
\begin{eqnarray}
\rho^{-1}\partial_i (\rho\partial^if_{IJ})
&=&f^{KL}(\partial^if_{IK})\partial_i f_{JL}
-f^{-1}(\partial^i\omega_I)\partial_i\omega_J,
\label{ellip_1}\\
\rho^{-1}\partial_i (\rho\partial^i\omega_I)
&=&f^{-1}(\partial^if)\partial_i\omega_I
+f^{JK}(\partial^if_{IJ})\partial_i\omega_K.
\label{ellip_2}
\end{eqnarray}
Once these equations are solved for five scalar functions $\{f_{IJ}\}$
and $\{\omega_I\}$, the other metric components are systematically found by integrations of
\begin{eqnarray}
\rho^{-1}\sigma_{,\rho}&=&
{1\over 8} f^{-2}[(f_{,\rho})^2-(f_{,z})^2]
+{1\over 8}f^{IJ}f^{KL}
(f_{IK,\rho}f_{JL,\rho}-f_{IK,z}f_{JL,z})
\nonumber\\
&&+{1\over 4}f^{-1}f^{IJ}
(\omega_{I,\rho}\omega_{J,\rho}-\omega_{I,z}\omega_{J,z}),
\nonumber\\
\rho^{-1}\sigma_{,z}&=&
{1\over 4} f^{-2}f_{,\rho}f_{,z}
+{1\over 4} f^{IJ}f^{KL} f_{IK,\rho} f_{JL,z}
+{1\over 2}f^{-1}f^{IJ}\omega_{I,\rho}\omega_{J,z},
\nonumber\\
w^I_{,\rho}&=&\rho f^{-1} f^{IJ}\omega_{J,z},
\nonumber\\
w^I_{,z}&=&-\rho f^{-1} f^{IJ} \omega_{J,\rho}.
\end{eqnarray}
The Equations (\ref{ellip_1}) and (\ref{ellip_2})
are derived via the extremization of the action
\begin{eqnarray}
S[\{f_{IJ}\},\{\omega_I\}]
&=&\int_{B^2} d\rho dz \rho\biggl[
{1\over 4}f^{-2}(\partial^i f)\partial_i f
+{1\over 4} f^{IJ} f^{KL}
(\partial^i f_{IK})\partial_i f_{JL}
\nonumber\\
&&+{1\over 2}f^{-1} f^{IJ}
(\partial^i\omega_I)\partial_i\omega_J
\biggr].\nonumber
\end{eqnarray}
This action principle is derived by Maison\cite{Maison79}. 
In this way, we now have a theory of five real scalar fields on $B^2$. 
Let us see how these five scalars can be
embedded in a matrix field in the following.
Define a symmetric real matrix field by
\begin{eqnarray}
\Phi =\left(
\begin{array}{ccc}
f^{-1}&-f^{-1}\omega_\phi&-f^{-1}\omega_{\psi}\\
\ast &f_{\phi\phi}+f^{-1}\omega_{\phi}\omega_{\phi}&f_{\phi\psi}+f^{-1}\omega_\phi\omega_{\psi}\\
\ast&\ast&f_{\psi\psi}+f^{-1}\omega_\psi\omega_\psi
\end{array}
\right).\nonumber
\end{eqnarray} 
The determinant of this matrix is unity at each point. 
Clearly, $\Phi$ is spectral decomposable at each point,
and all the eigenvalues 
are strictly positive in a point near the spatial infinity. 
Hence asymptotic flatness ensures that this matrix is strictly positive matrix everywhere.
We introduce the equivalence relation in $SL(3,\boldsymbol{R})$ by
\begin{eqnarray*}
  g\sim g'\Leftrightarrow \exists Q\in SO(3),~s.t.~g=g'Q.
\end{eqnarray*}
Then $\Phi$ uniquely determines $[\Phi^{1/2}]\in SL(3,\boldsymbol{R})/SO(3)$.
Conversely, let \\
$[g]\in SL(3,\boldsymbol{R})/SO(3)$ be given.
When the singular value decomposition of the representative $g$ is given by
\begin{eqnarray}
  g=PDQ^{-1},\nonumber
\end{eqnarray}
where $P$, $Q\in SO(3)$ and $D$ is the diagonal matrix composed of singular values of $g$,
\begin{eqnarray}
  \Phi=PD^2P^{-1}\nonumber
\end{eqnarray}
is clearly a strictly positive symmetric matrix with determinant 1, and
it is independent of the choice of the representative.
This shows that the matrix field $\Phi$ defines the differentiable map 
$B^2\to SL(3,\boldsymbol{R})/SO(3)$.

Let us define the current matrix field $\Pi_i$ by
\begin{eqnarray}
\Pi_i=\Phi^{-1}\Phi_{,i}.\nonumber
\end{eqnarray}
This belongs to the representation space of the adjoint representation
of global $SL(3,\boldsymbol{R})$ transformation.
In terms of this, the elliptic equations (\ref{ellip_1}) and (\ref{ellip_2}) are equivalent with
\begin{eqnarray}
\partial_i \Pi^i=0,\nonumber
\end{eqnarray}
and this is derived from the extremization of the
action
\begin{eqnarray}
S[\{\Pi_i\}]={1\over 4}\int_{B^2}d\rho dz \rho{\rm Tr}(\Pi^i\Pi_i).\nonumber
\end{eqnarray}
Hence, the problem is reduced to the (weighted) nonlinear $\sigma$-model over $B^2$ with the target space $SL(3,\boldsymbol{R})/SO(3)$.

Let us assume that there is a single non-degenerate
event horizon whose spatial section is diffeomorphic with a 3-sphere.
In this case, $B^2$ becomes $(\rho,z)$-half 2-plane given by $\rho\ge 0$.

The uniqueness theorem for the nonlinear $\sigma$-model on $B^2$ is obtained
utilising either the Mazur identity\cite{Mazur84}, 
or the Bunting identity\cite{Bunting83}.

The Mazur identity is in general applied for the nonlinear $\sigma$-model with the 
target space which is a coset space $G/H$.
Let both $\Phi$ and $\Phi'$ be possibly distinct solutions with the same asymptotic behavior.
The Mazur identity is derived from the divergence equation
\begin{eqnarray}
  \partial_i[\rho\partial^i{\rm Tr}(\Phi'\Phi^{-1}-I)]
=\rho{\rm Tr}[{}^t(\Pi^i{}'-\Pi^i)\Phi(\Pi_i'-\Pi_i)\Phi^{-1}],
\end{eqnarray}
where $I$ denotes the identity matrix field.
Integrating this equation over $B^2$, and applying the Green's theorem,
we obtain 
\begin{eqnarray}
 \oint_{\partial_{B^2}}\rho \partial^i{\rm Tr}(\Phi'\Phi^{-1}-I)] dS_i
=\int_{B^2}\rho {\rm Tr}({}^tM^iM_i)d\rho dz,
\label{Mazur}
\end{eqnarray}
where
\begin{eqnarray}
  M^i=g^{-1}{}^t(\Pi^i{}'-\Pi^i)g'
\end{eqnarray}
has been defined. The boundary conditions ensures that the left hand side of 
Eq.~(\ref{Mazur}) is zero. However, for the integrand in the right hand side 
of Eq.~(\ref{Mazur}) is nonnegative, it must be zero everywhere. 
This implies that $M^i=0$ holds everywhere. 
It immediately follows that two solutions $\Phi$ and $\Phi'$ are identical.

Another way to prove the uniqueness theorem is provided by the Bunting method 
in the following. 
\begin{eqnarray}
\partial_i[\rho\partial^i{\rm Dist}(x)^2]
=\rho G_{AB}h^{ij}(\mathscr{D}_i f^A_{,s})\mathscr{D}_jf^B_{,s}
-\rho{}^{\mathscr{T}^5}R_{ABCD}f^A_{,s}f^B_{,i}
f^C_{,s}f^D_{,i},
\label{Bunting_div}
\end{eqnarray}
where
\begin{eqnarray}
\mathscr{D}_{,i}f^A_{,s}=\partial_{i}f^A_{,s}
+{}^{\mathscr{T}^5}\Gamma^A_{BC}f^B_{,i}f^C_{,s}.
\nonumber
\end{eqnarray}
Since the first term on the right hand side of Eq.~(\ref{Bunting_div}) is nonnegative, we have
\begin{eqnarray}
\partial_i[\rho \partial^i{\rm Dist}(x)^2]\ge 0,
\end{eqnarray}
if the target space $\mathscr{T}^5$
  has nonpositive sectional curvature.
This is true in our case, as explained in the following. The Riemann curvature ${}^{\mathscr{T}^5}R^{AB}{}_{CD}$ defines a linear map acting on the 10-dimensional linear space of real antisymmetric tensors, by $V^{[AB]}\to {}^{\mathscr{T}^5}R^{AB}{}_{CD}V^{[CD]}$. This is a normal matrix and has only a triply degenerated nonzero eigenvalue $-5/2$. The nonpositivity of the sectional 
curvature of $\mathscr{T}^5$ immediately follows.

Integrating the divergence identity (\ref{Bunting_div}) over $B^2$ gives
\begin{eqnarray}
\int_{B^2}\partial_i\rho \partial^i{\rm Dist}(x)^2
d\rho dz
=\oint_{\partial_{B^2}}\rho \partial_i{\rm Dist}(x)^2 dS^i\nonumber.
\end{eqnarray}
If the integrand of the right hand side is zero, which, in the present case, 
is the case by the boundary conditions, the equality 
\begin{eqnarray}
\partial_i[\rho \partial^i{\rm Dist}(x)^2]=0
\end{eqnarray}
should hold. 
This implies that ${\rm Dist}(x)={\rm const.}$ 
However, this constant must be zero, for
the asymptotic conditions require ${\rm Dist(x)}\to 0$ ($\rho\to +\infty$). 
It immediately follows that $f^A(x)=f'{}^A(x)$ everywhere. 
Thus, we have shown the following theorem: 

\medskip 
\noindent 
{\bf Theorem~5.1:}~[Reference \citen{Morisawa:04}] 
{\em 
Consider $5$-dimensional asymptotically flat, stationary vacuum black holes 
with two commuting rotational Killing vector fields that commute also 
with the stationary Killing vector field. Assume further that the solutions 
have the spherical horizon topology, $S^3$. Then, the solution of this system 
can be uniquely determined by the three asymptotic charges, the mass $M$ and 
the two angular momenta, $J_1$ and $J_2$, and therefore is isometric to 
the Myers-Perry metric that has the same values of the corresponding 
asymptotic charges. 
} %

\medskip  
The above uniqueness result for spherical black holes has been extended 
to more general cases that include other horizon topologies by Hollands and 
Yazadjiev\cite{HollandsYazadjiev08} by employing the interval 
structure as a set of parameters to completely determine 
a black hole solution. We quote their theorem: 

\medskip 
\noindent 
{\bf Theorem~5.2:}~[Reference \citen{HollandsYazadjiev08}]  
{\em 
Consider in $5$-dimensions, two stationary, asymptotically flat, vacuum 
black hole spacetimes with non-degenerate horizon, 
having two commuting axial Killing fields that commute also with 
the stationary Killing field. 
Assume that the both solutions have the same interval structure and 
the same values of the angular momenta. Then they are isometric.  
} %

\medskip 
Thus, this theorem encompasses black ring solutions\cite{ER02b,PS06} 
as well as hypothetical black lens solutions\footnote{ 
Some attempts to find an exact black lens solution 
have been made in Ref.~\citen{Evslin08} [see also, e.g.,
Ref.~\citen{ChenTeo08}].
However, the `black lens solutions' so far constructed turned out
to suffer from a naked singularity. 
} 
(if such a solution exists), besides the spherical black holes\cite{MP86}.
Note that the mass can be determined by the angular momenta and
the interval structure. 
It is worth emphasising that the interval structure 
can completely determine not only the topology of each connected component 
of the event horizon but also the topology of the domain of outer 
communications, as well as the action of the rotational isometries. 
The above theorem has been further generalized to include asymptotically 
Kaluza-Klein black holes\cite{HollandsYazadjiev08b}.

\section{Summary}
\label{sec:summary} 
We have seen basic properties of black holes in higher dimensions, 
focusing mainly on the subjects relevant to the uniqueness theorems for 
asymptotically flat, stationary vacuum solutions in general relativity. 
After a brief summary of the global aspects relevant to the black hole 
area theorem in higher dimensions, we have seen rich topological aspects of 
dynamical, non-stationary black holes\cite{Siino98,IdaSiino07}
in section~\ref{sec:general}, where 
the crease set plays a key role. Galloway and Schoen's topology 
theorem\cite{GallowaySchoen06} provides the non-trivial restriction on 
possible topology of apparent as well as the event horizons. 
When a black hole spacetime admits axial/rotational symmetries, we have 
topological restrictions stronger than that of Galloway and Schoen as seen in 
section~\ref{subsec:topology}. 
This is relevant in particular when we are concerned with stationary 
black holes since `stationary' implies `axisymmetric' due to 
the rigidity theorem. 
In particular, when a stationary black hole admits multiple rotational 
symmetries so that the factor space of the black hole spacetime 
by its isometry reduces to $2$-dimensional, the rod/interval 
structure\cite{Harmark04,HollandsYazadjiev08} can completely specify 
the horizon topology as briefly commented in section~\ref{subsec:topology} 
and at the end of section~\ref{subsec:unique:stationary:5D}. 
When there is only a single rotational symmetry in $5$-dimensions, 
we have a weaker topological restriction\cite{HHI10}, which allows for 
the possibility of `Black Prism' and other various Seifert manifolds 
as the horizon cross section manifold, 
though such a larger variety of the horizon manifold than those already 
known in the existing exact solutions may possibly be 
ruled out once the Einstein equations are fully taken into account.  
In section~\ref{subsec:rigidity}, we have seen the main idea of 
the rigidity proof in higher dimensions\cite{HIW07}, 
and its extension to extremal black holes with some additional 
technical assumption\cite{HI09}. It guarantees the existence of only 
one rotational symmetry, which is in accord with the conjecture of 
existence of less symmetric stationary black holes\cite{Reall03}, 
while all known exact solutions have multiple rotational symmetries. 
For this reason, in order to obtain some uniqueness results by following 
basic steps described in 
section~\ref{subsec:unique:stationary:strategy}---similar to Steps (i)--(iv) 
of the $4$-dimensional case, one has to impose multiple $(D-3)$-axial 
symmetries, besides the assumed stationary symmetry. By doing so, 
one can prove the uniqueness of stationary rotating black holes in higher 
dimensions with some additional data, such as a spherical 
topology\cite{Morisawa:04} or more generally the rod/interval 
structure\cite{HollandsYazadjiev08}, other than the asymptotic conserved 
charges as seen in section~\ref{subsec:unique:stationary:5D}. 
For the static case, the existing proof in $4$-dimensions uses some geometric 
properties that hold only in $4$-dimensions, and therefore it is not 
straightforward to generalize to higher dimensions. We have 
seen how to overcome this difficulty\cite{Hwang98,Gibbons02b} in 
section~\ref{sec:unique:static}.

The uniqueness results in higher dimensions discussed in 
section~\ref{sec:unique:stationary} hold only for non-extremal black holes. 
In $4$-dimensions, generalizations to 
uniqueness theorems for extremal Kerr and extremal charged Kerr black holes 
have recently been made by Refs.~\citen{CN10,AHMR10,FL09}. 
The key new element for the proof is the uniqueness of the near-horizon 
geometry for a degenerate horizon of stationary axisymmetric vacuum 
spacetimes\cite{Hajicek74,LP03,KL08}, i.e., the near-horizon geometry 
must agree with that of the extremal Kerr metric\cite{BardeenHorowitz99},  
and a similar result also holds for the extremal electrovacuum black 
hole case.  
In $5$-dimensions, however, near horizon geometries for stationary 
vacuum solutions with two rotational symmetries are not unique. 
[See Ref.~\citen{KL08} for the lack of the uniqueness and 
Refs.~\citen{KL08,KL09b,FKLR08,KLR07,HI10} for classification of 
near-horizon geometries.] Therefore it seems to be 
highly non-trivial whether one can obtain similar uniqueness results 
for extremal black holes in higher dimensions. 
In this regard, it is interesting to note Theorem~2 of Ref.~\citen{FL09} 
that the interval structure can be used to uniquely determine extremal vacuum 
black holes in $5$-dimensions, (as well as near-horizon geometries) 
under similar assumptions of the uniqueness 
theorem~5.2 above\cite{HollandsYazadjiev08}.

It turns out that on one hand, the rod/interval structure provides 
a convenient set of data to uniquely determine a stationary 
black hole spacetime with $(D-3)$ rotational symmetries. 
It may be possible to completely classify all stationary vacuum 
black hole solutions, using the rod/interval structure. 
On the other hand, as it is a local notion, the use of the rod/interval 
structure in uniqueness theorems does not appear to be fully satisfactory; 
an asymptotically flat black hole, as an isolated system, should be described 
in terms of asymptotic data as accurate as possible. 
However, as indicated by the existence of multiple black objects such as 
black Saturn\cite{Elvang&Figueras07} and di-ring\cite{IguchiMishima07}, 
if attempting to completely characterize a black hole solution 
by using asymptotic data (e.g., multipole moments defined at infinity) 
alone, then one would need to identify an infinite set of 
the multipole moments.  
In view of this, it would be interesting to clarify relations between 
the rod structure and asymptotic conserved charges and multipole moments. 
[See Ref.~\citen{TOS10} for recent progress along this line.]  
The task would be to identify a minimal set of local data (e.g., the number 
of connected components of the event horizon) and a maximal finite set of 
asymptotic data.

\section*{Acknowledgments}
AI is supported by the Grant-in-Aid for Scientific Research Fund of the JSPS 
(A)No. 22244030 (C)No. 22540299.
TS is supported by the Grant-in-Aid for Scientific Research Fund of the JSPS 
No. 21244033, No. 21111006, No. 20540258 and 19GS0219.

%

\end{document}